\documentclass[11pt,ams]{article}
\usepackage{graphicx}
\usepackage{pslatex}
\usepackage{amsmath}
\usepackage{amssymb}
\usepackage{a4wide}

\newcommand{\sect}[1]{ \section{#1} \setcounter{equation}{0} }

\begin{document}
\date{}
\title{\textbf{The Gribov-Zwanziger action in the presence of the gauge invariant, nonlocal mass operator $\mathrm{Tr} \int d^4x
F_{\mu\nu} (D^2)^{-1} F_{\mu\nu}$ in the Landau gauge}}
\author{ \textbf{M.A.L. Capri$^a$\thanks{marcio@dft.if.uerj.br}} \ , \textbf{D. Dudal}$^{b}$\thanks{david.dudal@ugent.be} \ ,
\textbf{V.E.R. Lemes$^{a}$\thanks{vitor@dft.if.uerj.br}}  \ , \\\textbf{R.F. Sobreiro}$^{c}$\thanks{%
sobreiro@cbpf.br}  \ , \textbf{S.P. Sorella}$^{a}$\thanks{%
sorella@uerj.br}{\ } \ , \textbf{R.
Thibes$^{a}$\thanks{thibes@dft.if.uerj.br}}
\ ,  \textbf{H. Verschelde}$^{b}$\thanks{henri.verschelde@ugent.be} \\\\
\textit{$^{a}$\small{Departamento de F\'{\i }sica Te\'{o}rica, Instituto de F\'{\i }sica, UERJ, Universidade do Estado do Rio de Janeiro}}\\
\textit{\small{Rua S{\~a}o Francisco Xavier 524, 20550-013
Maracan{\~a}, Rio de Janeiro, Brasil}} \\ [2mm]
\textit{$^{b}$\small{Ghent University, Department of Mathematical Physics and Astronomy}} \\
\textit{\small{Krijgslaan 281-S9, B-9000 Gent, Belgium}}\\ [2mm]
\textit{$^{c}$\small{CBPF, Centro Brasileiro de Pesquisas
F{\'\i}sicas}} \\
\textit{\small{ Rua Xavier Sigaud 150, 22290-180, Urca, Rio de
Janeiro, Brasil}} } \maketitle \vspace{-20pt}
\begin{abstract}
\noindent We prove that the nonlocal gauge invariant mass dimension
two operator $F_{\mu\nu} (D^2)^{-1} F_{\mu\nu}$ can be consistently
added to the Gribov-Zwanziger action, which implements the
restriction of the path integral's domain of integration to the
first Gribov region when the Landau gauge is considered. We identify
a local polynomial action and prove the renormalizability to all
orders of perturbation theory by employing the algebraic
renormalization formalism. Furthermore, we also pay attention to the
breaking of the BRST invariance, and to the consequences that this
has for the Slavnov-Taylor identity.
\end{abstract}
\newpage

\sect{Introduction}

It is a well known fact that SU(N) Yang-Mills gauge theories, described by
the Euclidean action
\begin{equation}
S_{\mathrm{YM}}=\frac{1}{4}\int d^{4}\!x\,F_{\mu \nu }^{a}F_{\mu \nu }^{a}\;,
\label{1}
\end{equation}
with $A_{\mu }$ the gauge potential and
\begin{equation}
F_{\mu \nu }^{a}=\partial _{\mu }A_{\nu }^{a}-\partial _{\nu }A_{\mu
}^{a}+gf^{abc}A_{\mu }^{b}A_{\nu }^{c}\;,
\end{equation}
the field strength, whereby $D_{\mu }^{ab}$ is the covariant derivative in
the adjoint representation, given by
\begin{equation}
D_{\mu }^{ab}=\partial _{\mu }\delta ^{ab}-gf^{abc}A_{\mu }^{c}\;,
\end{equation}
are asymptotically free at very high energies \cite
{Gross:1973id,Politzer:1973fx,Jones:1974mm,Caswell:1974gg}. The coupling
constant is sufficiently small to allow for a perturbative description, with
asymptotic degrees of freedom given by massless gauge bosons. We shall not
consider fermion matter in this paper, however the same conclusion holds for
quantum chromodynamics (QCD), written in terms of gluons and quarks. When we
pass to lower energies, the coupling constant $g^{2}$ begins to grow, and
perturbation theory starts to loose its validity. At still lower energies,
the situation becomes dramatically different, as perturbation theory now
completely fails, and confinement sets in, meaning that the elementary field
excitations are no longer physical observables, but become confined into
colorless states. The hadrons constitute the physical states of QCD.

\noindent A satisfactory understanding of the behaviour of Yang-Mills
theories in the low energy regime is yet unavailable. Due to the large
coupling constant, nonperturbative effects have to be taken into account.
The introduction of condensates, which are the (integrated) vacuum
expectation value of certain local operators, allows one to parametrize
certain nonperturbative effects arising from the infrared sector of e.g. the
theory described by (\ref{1}). Condensates give rise to power corrections, a
phenomenon that can be handled using the Operator Product Expansion.
Clearly, these power corrections correspond to nonperturbative information
in addition to the perturbatively calculable results. If one wants to
consider the possible effects of condensates on physical quantities in a
gauge theory, only gauge invariant operators are relevant. The most famous
example is the dimension 4 gluon condensate $\left\langle\alpha_s
F_{\mu\nu}^2\right\rangle$, giving rise to $\frac{1}{q^4}$ power corrections
in QCD. The SVZ (Shifman-Vainshtein-Zakharov) sum rules \cite{Shifman:1978bx}
can then be used to relate the condensates to observables, and hence one may
obtain certain phenomenological estimates for e.g. $\left\langle\alpha_s
F_{\mu\nu}^2\right\rangle$. This approach allows for a study of at least
some aspects of QCD in an energy regime in between the confined and
perturbative zone. One can still use perturbation theory using the quarks
and gluons as effective degrees of freedom due to the lack of explicit
knowledge of the correct physical degrees of freedom, but the results get
adapted by nonperturbatively generated condensates.

\noindent In this paper, we shall introduce an action that can serve as a
starting point for investigating some nonperturbative effects in gauge
theories. These nonperturbative effects arise from 2 premises: the Gribov
problem in fixing the gauge freedom, and the possibility of a dynamical mass
generation.

\noindent First of all, we shall be concerned about fixing the gauge. If we
want to perform any kind of calculations, we must reduce the enormous gauge
freedom of (\ref{1}), encoded in the local gauge symmetry generated by
\begin{equation}
\delta _{\omega }A_{\mu }^{a}=-D_{\mu }^{ab}\omega ^{b}\,,\qquad \mathrm{%
with\;}\omega ^{b}\mbox{
    arbitrary}\;,  \label{gaugefreedom}
\end{equation}
to a global one by a suitable gauge fixing condition, say $\mathcal{F}%
(A_{\mu })=0$. In principle, by imposing a gauge condition one should select
a single representative $A_{\mu }^{*}$ from the gauge orbit $%
A_{\mu }^{U}$, where $U$ is a generic $SU(N)$ gauge transformation.
Unfortunately, it was shown that is impossible to uniquely fix the gauge, a
problem related to the complicated topology of the space of gauge orbits
\cite{Singer:1978dk}.

\noindent Seminal work on the existence of gauge copies was done 3
decades ago by Gribov in \cite{Gribov:1977wm}. This paper is not the
place to give a complete overview of the ambiguities arising when a
gauge fixing is performed, we therefore kindly refer to Gribov's
original paper or to the available literature, such as
\cite{Sobreiro:2005ec}, which contains many examples and references.
In particular, in \cite{Gribov:1977wm}, it was pointed out that, in
the Landau and Coulomb gauges, the existence of zero modes in the
Faddeev-Popov operator gives rise to gauge copies. Using the Landau
gauge,
\begin{equation}
\partial _{\mu }A_{\mu }=0\;,  \label{landau}
\end{equation}
one finds that a gauge equivalent configuration $A_{\mu }^{\prime }$,
connected to $A_{\mu }$ via \eqref{gaugefreedom}, also obeys $\partial _{\mu
}A_{\mu }^{\prime }=0$, when
\begin{equation}
\mathcal{M}^{ab}\omega ^{b}=0\;,  \label{landau2}
\end{equation}
where $\mathcal{M}^{ab}$ denotes the Faddeev-Popov operator
\begin{equation}
\mathcal{M}^{ab}=-\partial _{\mu }(\partial _{\mu }\delta
^{ab}-gf^{abc}A_{\mu }^{c})\;.  \label{FPop}
\end{equation}

\noindent The existence of the Gribov copies implies that the domain of
integration in the path integral has to be further restricted in a suitable
way. Following Gribov, it seems logical to restrict to the region $\Omega $
with corresponding boundary $\partial \Omega $, which is the first Gribov
horizon, where the first vanishing eigenvalue of the Faddeev-Popov operator %
\eqref{FPop} appears \cite{Gribov:1977wm}. Within the region $\Omega
$ the Faddeev-Popov operator is positive definite, i.e.
$\mathcal{M}^{ab}>0$. Quite obviously, this restriction to the first
Gribov region can be motivated only if every gauge orbit passes
through it. It was shown by Gribov that this is certainly the case
for gauge potentials ``sufficiently close'' to the boundary
$\partial \Omega $ \cite{Gribov:1977wm}, whereas the
proof for general configurations was presented in \cite{Dell'Antonio:1991xt}%
. Nevertheless, we should also mention that the Gribov region itself
is also not free from gauge copies \cite
{Dell'Antonio:1991xt,Semenov,Dell'Antonio2,vanBaal:1991zw}. To avoid
the presence of these additional copies, a further restriction to a
smaller region $\Lambda $, known as the fundamental modular region,
should be implemented. Nevertheless, the implementation of the
restriction of the domain of integration to $\Lambda $ proves to be
a quite difficult task which, to our knowledge, has not yet been
accomplished.
Recently, it has been argued that the additional copies existing inside $%
\Omega $ might be irrelevant when computing expectation values, meaning that
averages calculated over $\Lambda $ or $\Omega $ should give the same value
\cite{Zwanziger:2003cf}.

\noindent Using a semiclassical argument \cite{Gribov:1977wm},
Gribov implemented the restriction to the region $\Omega $.
Essentially, his argument relied on the fact that the (Fourier
transform of the) inverse of the Faddeev-Popov operator, which is
nothing else than the ghost propagator, encounters no poles
elsewhere than at the origin $k^{2}=0$. This amounts to say that the
operator $\mathcal{M}^{ab}$ itself does not vanish, except at the
horizon. By using this ``no pole condition'', we are assured that
the considered gauge potentials remain inside the first Gribov
region\footnote{%
Albeit that this restriction cannot be implemented exactly, but only in an
order by order expansive way.}, and as a such at least the set of copies
obtained via \eqref{landau2} is already excluded from the game.

\noindent This restriction has many important consequences for the
infrared behaviour of the propagators. The gluon propagator turns
out to be suppressed in the infrared, while the ghost propagator
gets enhanced \cite {Gribov:1977wm}. Moreover, it can also be shown
that the gluon propagator exhibits a violation of positivity in its
spectral density representation, a sign that the gluon cannot be a
physical observable anymore, see \cite
{Dudal:2005na,Alkofer:2003jj,Cucchieri:2004mf,Bowman:2007du} and
references therein. It is interesting to mention that lattice
simulations of the Landau gauge propagators have revealed evidence
for this suppression, respectively enhancement, see e.g. \cite
{Cucchieri:1999sz,Bonnet:2001uh,Langfeld:2001cz,Cucchieri:2003di,Bloch:2003sk,Furui:2003jr,Furui:2004cx,Suman:1995zg,Bakeev:2003rr}%
. Another consequence of the Gribov restriction is the ``infrared
freezing'' of the strong coupling constant, i.e. $\alpha
_{s}(p^{2})$ tends to a constant as $p^{2}$ goes to zero, see
\cite{Dudal:2005na,Gracey:2006dr} and references therein. Again,
this behaviour is in qualitative agreement with lattice data
\cite{Bloch:2003sk,Furui:2003jr,Furui:2004cx} as well as with the
results obtained from the analysis of the Schwinger-Dyson equations
\cite
{vonSmekal:1997is,vonSmekal:1997is2,Atkinson:1997tu,Atkinson:1998zc,Alkofer:2000wg,Watson:2001yv,Zwanziger:2001kw,Lerche:2002ep,Fedosenko:2007tj}%
.

\noindent It might be clear that the restriction to the Gribov
region $\Omega $ could be of great relevance for a better
understanding of the infrared region of gauge theories. This belief
is further supported by the Kugo-Ojima confinement criterion
\cite{Kugo:1979gm} which, in the case of the Landau gauge, turns out
to rely on a ghost propagator diverging stronger than
$\frac{1}{p^{2}}$ \cite{Kugo:1995km}. This feature is also present
when the restriction to the Gribov region is
implemented, yielding in fact a ghost propagator developing a $\frac{1}{p^{4}%
}$ singularity.

\noindent An important progress on the restriction to the Gribov
region $\Omega $ was accomplished by Zwanziger in the papers \cite
{Zwanziger:1989mf,Zwanziger:1992qr}. The restriction to $\Omega $
was implemented through the introduction of a nonlocal horizon
function appearing in the Boltzmann weight defining the Euclidean
Yang-Mills measure. According to
\cite{Zwanziger:1989mf,Zwanziger:1992qr}, the starting Yang-Mills
measure in the Landau gauge is given by
\begin{equation}
d\mu _{\gamma }=DADcD\overline{c}Dbe^{-\left( S+\gamma ^{4}H\right) }\;,
\label{m1}
\end{equation}
where the starting action is
\begin{equation}
S=S_{\mathrm{YM}}+S_{\mathrm{gf}}\;,
\end{equation}
with $S_{\mathrm{gf}}$ the gauge-fixing action given by
\begin{equation}
S_{\mathrm{gf}}=\int d^{4}\!x\,\left( b^{a}\partial _{\mu }A_{\mu }^{a}+%
\overline{c}^{a}\text{ }\partial _{\mu }D_{\mu }^{ab}c^{b}\right)
\;,  \label{sgf}
\end{equation}
where the auxiliary field $b^{a}$ is a Lagrange multiplier enforcing
the Landau gauge \eqref{landau}, $(\overline{c}^{a},c^{a})$ are the
Faddeev-Popov ghost fields, and
\begin{equation}
H=\int d^{4}xh(x)=g^{2}\int d^{4}xf^{abc}A_{\mu }^{b}\left( \mathcal{M}%
^{-1}\right) ^{ad}f^{dec}A_{\mu }^{e}\;,  \label{m3}
\end{equation}
is the so-called horizon function, which implements the restriction
to the Gribov region $\Omega $. We recognize that $H$ is nonlocal.
The massive Gribov parameter $\gamma $ is fixed by the horizon
condition
\begin{equation}
\left\langle h(x)\right\rangle =4\left( N^{2}-1\right) \;,  \label{m4}
\end{equation}
where the expectation value $\left\langle h(x)\right\rangle $ has to be
evaluated with the measure \eqref{m1}. To the first order, the horizon
condition (\ref{m4}) becomes, in $d$ dimensions,
\begin{equation}
1=\frac{N\left( d-1\right) }{4}g^{2}\int \frac{d^{d}q}{\left( 2\pi \right)
^{d}}\frac{1}{q^{4}+2Ng^{2}\gamma ^{4}}\;.  \label{mm5}
\end{equation}
This equation coincides with the original gap equation derived by Gribov for
the parameter $\gamma $ \cite{Gribov:1977wm}.

\noindent We shall rely on the path integral formalism, so that we can
localize the horizon fuction \eqref{m3} by means of a pair of complex
bosonic vector fields \cite{Zwanziger:1992qr}, $(\phi _{\mu }^{ab},\overline{%
\phi }_{\mu }^{ab})$, according to
\begin{equation}
e^{-S_{\mathrm{H}}}=\int D\overline{\phi }D\phi \,\,(\det \mathcal{M}%
)^{f}\exp \left\{ -\int d^{4}\!x\,\left[ \overline{\phi }_{\mu }^{ac}%
\mathcal{M}^{ab}\phi _{\mu }^{bc}+\gamma ^{2}gf^{abc}(\phi _{\mu }^{ac}-%
\overline{\phi }_{\mu }^{ac})A_{\mu }^{b}\right] \right\} \;,
\end{equation}
where the determinant, $(\det \mathcal{M})^{f}$, takes into account the
Jacobian arising from the integration over $(\phi _{\mu }^{ab},%
\overline{\phi }_{\mu }^{ab})$, and
\begin{equation}
f=D(N^{2}-1)=4(N^{2}-1)\;,
\end{equation}
with $D=4$ the dimension of the Euclidean space time, and $N$ the dimension
of the gauge group. This determinant can also be localized by means of
suitable anticommuting complex vector fields $(\omega _{\mu }^{ab},\overline{%
\omega }_{\mu }^{ab})$, namely
\begin{equation}
(\det \mathcal{M})^{f}=\int D\overline{\omega }D\omega \,\exp \left[ -\int
d^{4}\!x\,\left( -\overline{\omega }^{ac}\mathcal{M}^{ab}\omega _{\mu
}^{bc}\right) \right] \;.
\end{equation}
Henceforth, the nonlocal action $S_{\mathrm{H}}$ is transformed into a local
one given by
\begin{equation}
S_{\mathrm{H}}^{\mathrm{Local}}=S_{\phi \omega }+S_{\gamma }\;,
\label{local-H}
\end{equation}
where
\begin{equation}
S_{\phi \omega }=\int d^{4}\!x\,\left( \overline{\phi }_{\mu }^{ac}\mathcal{M%
}^{ab}\phi _{\mu }^{bc}-\overline{\omega }_{\mu }^{ac}\mathcal{M}^{ab}\omega
_{\mu }^{bc}\right) \;,
\end{equation}
and
\begin{equation}
S_{\gamma }=\gamma ^{2}\int d^{4}\!x\,gf^{abc}(\phi _{\mu }^{ac}-\overline{%
\phi }_{\mu }^{ac})A_{\mu }^{b}\;.  \label{Sgamma}
\end{equation}
As it was shown in \cite
{Dudal:2005na,Zwanziger:1989mf,Zwanziger:1992qr,Maggiore:1993wq},
the resulting local action turns out to be renormalizable to all
orders of perturbation theory. This is a point of great importance,
as it allows for a consistent and order by order improvable
framework to calculate relevant quantities when the restriction to
the Gribov region $\Omega $ is taken into account.

\noindent A second point that motivated this paper is the issue of
the dynamical mass generation in gauge theories, and related to it
that of the $\frac{1}{q^{2}}$ power corrections. A few years ago, in
a series of papers, Zakharov et al. questioned the common wisdom
that $\frac{1}{q^{2}}$ power corrections cannot enter gauge
invariant observables, as local gauge invariant operators of mass
dimension two do not exist. This is a reflection of the fact that
one cannot add a renormalizable mass operator for the gauge fields
to the Yang-Mills action, at least not when the Higgs mechanism and
associated symmetry breaking are not considered. However, by using
QCD sum rules, it was advocated in \cite{Chetyrkin:1998yr} that an
effective gluon mass could account for the $\frac{1}{q^{2}}$
corrections, leading to an acceptable phenomenology. The underlying
condensate was proposed to be the gauge invariant quantity \cite
{Gubarev:2000eu,Gubarev:2000nz}
\begin{equation}
\left\langle A_{\min }^{2}\right\rangle \equiv \min_{U\in SU(N)}\frac{1}{VT}%
\int d^{4}x\left\langle \left( A_{\mu }^{U}\right) ^{2}\right\rangle \;,
\label{intro1}
\end{equation}
which originates from a highly nonlocal operator, since
\cite[and references therein]{Capri:2005dy}
\[
A_{\min }^{2}=\frac{1}{2}\int d^{4}x\left[ A_{\mu }^{a}\left( \delta _{\mu
\nu }-\frac{\partial _{\mu }\partial _{\nu }}{\partial ^{2}}\right) A_{\nu
}^{a}-gf^{abc}\left( \frac{\partial _{\nu }}{\partial ^{2}}\partial
A^{a}\right) \left( \frac{1}{\partial ^{2}}\partial {A}^{b}\right) A_{\nu
}^{c}\right] \;+O(A^{4})\;,
\]
and therefore it falls beyond the OPE applicability. The interest
was especially focused on the Landau gauge, since then the operator
$A_{\min }^{2}$ reduces to the local quantity $A_{\mu }^{2}$. An
effective potential for $\langle A_{\mu }^{2}\rangle $ was
calculated up to two loops in
\cite{Verschelde:2001ia,Browne:2003uv}, giving evidence for a
nonvanishing condensate and consequent effective gluon mass
$m^{2}\propto \langle A_{\mu }^{2}\rangle $. Determining a sensible
effective potential for a local composite operator (LCO) is a
nontrivial task, but nevertheless it was dealt with in
\cite{Verschelde:2001ia} based on the method developed in
\cite{Verschelde:1995jj}. The renormalizability of the so-called LCO
method was proven in \cite{Dudal:2002pq} to all orders of
perturbation theory in the case of $A_{\mu }^{2}$ in the Landau
gauge.

\noindent Unfortunately, the gauge invariance of $A_{\min }^{2}$ is, strictly speaking, only ensured when the \emph{%
absolute} minimum of $A_{\min }^{2}$ along the gauge orbit has been
reached, a highly difficult task due to the presence of Gribov
copies. Moreover, it is unclear what can be done with this operator
beyond the Landau gauge. In other gauges, other renormalizable and
condensing dimension two operators exist, but these are explicite
gauge parameter or even ghost dependent, see \cite{Dudal:2006xd} for
an overview. In e.g. the maximal Abelian gauge, an effective mass
was found for the off-diagonal gluons only \cite{Dudal:2004rx},
which is qualitatively consistent with the available lattice data
\cite{Amemiya:1998jz,Bornyakov:2003ee}.

\noindent Let us also mention that effective gluon masses have been
studied in the past from theoretical, phenomenological and numerical
viewpoint, see \cite
{Langfeld:2001cz,Verschelde:2001ia,Parisi:1980jy,Field:2001iu,Bernard:1981pg,Marenzoni:1994ap,
Cornwall:1981zr,Aguilar:2004sw,Aguilar:2006gr,Halzen:1992vd} for a
far from exhaustive list.

\noindent Taking all this into account, it seems to be a worthy task
to look for other potential candidates which could be at the origin
of the dynamical mass generation and related $\frac{1}{q^{2}}$ power
corrections. It would also be favourable to start from a gauge
invariant operator. The candidate we already investigated in \cite
{Capri:2005dy,Capri:2006ne} is the nonlocal operator
\begin{equation}
\mathcal{O}=(VT)^{-1}\int d^{4}\!x\,F_{\mu \nu }^{a}\left[ \left(
D^{2}\right) ^{-1}\right] ^{ab}F_{\mu \nu }^{b}\;,  \label{mass-op}
\end{equation}
which can be coupled to the Yang-Mills action via a nonlocal mass term
\begin{equation}
S_{\mathcal{O}}=-\frac{m^{2}}{4}\int d^{4}\!x\,F_{\mu \nu }^{a}\left[ \left(
D^{2}\right) ^{-1}\right] ^{ab}F_{\mu \nu }^{b}\;,  \label{mass-op}
\end{equation}
where $m$ is a mass parameter, and $[(D^{2})^{-1}]^{ab}$ is the inverse of
the covariant Laplacian
\begin{equation}
D^{2}\equiv D_{\mu }^{ac}D_{\mu }^{cb}=\partial ^{2}\delta
^{ab}-2gf^{abc}A_{\mu }^{c}\partial _{\mu }-gf^{abc}\partial _{\mu }A_{\mu
}^{c}+g^{2}f^{acd}f^{cbe}A_{\mu }^{d}A_{\mu }^{e}\;.
\end{equation}
Analogously to what has been done in the case of the nonlocal
horizon function \eqref{m3}, the gauge invariant mass operator
\eqref{mass-op} can be localized with the help of a pair of complex
bosonic
antisymmetric tensor fields in the adjoint representation \cite{Capri:2005dy}%
, $(B_{\mu \nu }^{a},\overline{B}_{\mu \nu }^{a})$,
\begin{equation}
e^{-S_{\mathcal{O}}}=\int D\overline{B}DB\,(\det D^{2})^{f^{\prime }}\,\exp
\left\{ -\frac{1}{4}\int d^{4}\!x\,\left[ \overline{B}_{\mu \nu
}^{a}D_{\sigma }^{ac}D_{\sigma }^{cb}B_{\mu \nu }^{b}+im(B_{\mu \nu }^{a}-%
\overline{B}_{\mu \nu }^{a})F_{\mu \nu }^{a}\right] \right\} \;.
\end{equation}
Here we have
\begin{equation}
f^{\prime }=\frac{D(D-1)}{2}=6\;,
\end{equation}
and, like in the case of the horizon function, the determinant, $(\det
D^{2})^{f^{\prime }}$, can be localized using a pair of anticommuting
antisymmetric complex tensor fields $(G_{\mu \nu }^{a},\overline{G}_{\mu \nu
}^{a})$, according to
\begin{equation}
(\det D^{2})^{f^{\prime }}=\int D\overline{G}DG\,\exp \left[ -\frac{1}{4}%
\int d^{4}\!x\,\left( \overline{G}_{\mu \nu }^{a}D_{\sigma }^{ac}D_{\sigma
}^{cb}G_{\mu \nu }^{b}\right) \right] \;.
\end{equation}
Then, the action $S_{\mathcal{O}}$ gets replaced by its local version given
by
\begin{equation}
S_{\mathcal{O}}^{\mathrm{Local}}=S_{BG}+S_{m}\;,  \label{Llocal-O}
\end{equation}
where,
\begin{equation}
S_{BG}=\frac{1}{4}\int d^{4}\!x\,\left( \overline{B}_{\mu \nu }^{a}D_{\sigma
}^{ac}D_{\sigma }^{cb}B_{\mu \nu }^{b}-\overline{G}_{\mu \nu }^{a}D_{\sigma
}^{ac}D_{\sigma }^{cb}G_{\mu \nu }^{b}\right) \;,
\end{equation}
and
\begin{equation}
S_{m}=\frac{im}{4}\int d^{4}\!x\,(B_{\mu \nu }^{a}-\overline{B}_{\mu \nu
}^{a})F_{\mu \nu }^{a}\;.  \label{Sm}
\end{equation}
We underline the fact that an initially nonlocal operator can be
cast into a local form \cite{Capri:2007ck}. In the case of $A_{\min
}^{2}$, this would not be possible, as it is a infinite series of
different nonlocal operators. Once we arrive at a local action, we
can investigate e.g. the renormalizability to all orders by means of
algebraic methods, the canonical quantization, the explicit
calculation of the renormalization factors, etc.

\noindent The goal of this paper is to study the massive action
\eqref{mass-op} when the restriction to the Gribov region $\Omega $
is
implemented \`{a} la Zwanziger. Since the extended action $S_{YM}+S_{%
\mathcal{O}}$ is gauge invariant, we might expect that the procedure
of further restricting the domain of integration will have no
influence on the renormalizability. This will be explicitly
confirmed. In a future stage of research, one can start searching
for the value of the Gribov parameter $\gamma $ as well as the
dynamically generated mass $m$. We remind here that the
Gribov-Zwanziger action itself can also be used to mimic
$\frac{1}{q^{2}}$ corrections, as explicitly discussed in
\cite{Gracey:2006dr}. As a future endeavour, it would be worthwhile
to study physical correlators with our action, and find whether the
Gribov and/or mass parameter $m$ are a potential source of such
power corrections.

\noindent Let us return to the content of this paper, which is
organized as follows. In section 2, we introduce all the necessary
sources in order to find a suitable starting action. The set of Ward
identities defining this action is presented in section 3, while in
section 4 we compute several useful (anti-)commutation relations
between the linearized symmetry operators. These are used in section
5 in order to construct the most general allowed invariant
counterterm. In section 6, we confirm the renormalizability since we
shall be able to reabsorb all the allowed counterterms in the
starting action by introducing suitable bare quantities. In section
7 we discuss a few properties of the physical action, which is
obtained from the starting action by setting the sources equal to
their physically relevant values. The process of giving specific
values to the sources breaks the BRST invariance. In section 8, we
discuss the associated breaking of the Slavnov-Taylor identity, and
we comment on the fact that in most cases this breaking becomes
harmless for the identities derivable between a large class of Green
functions. Finally, section 9 is devoted to the conclusions.

\sect{Identification of the complete classical action $\Sigma$}
\noindent We shall start with the following local action, as it was
obtained in the introduction
\begin{equation}
S_{\mathrm{Local}}=S_{\mathrm{YM}}+S_{\mathrm{gf}}+S_{\mathrm{H}}^{\mathrm{%
Local}}+S_{\mathcal{O}}^{\mathrm{Local}}\;.  \label{local}
\end{equation}
As we wish to discuss the renormalizability, we should try to
establish as many symmetries as possible. These symmetries can then
be translated into Ward identities. As we are dealing with a gauge
theory which is to be gauge fixed, we expect to find a BRST
invariance and consequent Slavnov-Taylor identity. All these
identities are a powerful tool in constructing the most general
allowed counterterm \cite {Piguet:1995er}. If this counterterm can
be reabsorbed in the original action through the introduction of
bare quantities, we are able to conclude that the starting action is
renormalizable. If not, we could still try to identify a more
general starting action that is renormalizable. This has been
discussed in extenso already in \cite{Capri:2005dy,Capri:2006ne}
when analyzing the nonlocal mass term \eqref{mass-op}.

\subsection{BRST invariance}

In order to find the BRST invariance of the resulting local theory, given by %
\eqref{local}, we proceed as in \cite{Zwanziger:1992qr,Capri:2005dy} and
consider at first the particular case when $\gamma =m=0$, i.e.,
\begin{eqnarray}
S_{\mathrm{Local}}^{\gamma =m=0} &=&S_{\mathrm{YM}}+S_{\mathrm{gf}}+S_{%
\mathrm{H}}^{\mathrm{Local},\gamma =0}+S_{\mathcal{O}}^{\mathrm{Local},m=0}
\nonumber \\
&=&S_{\mathrm{YM}}+S_{\mathrm{gf}}+S_{\phi \omega }+S_{BG}\;.\phantom{\Bigl|}
\label{local-zero}
\end{eqnarray}
In this case, we have actually introduced nothing more than two
unity factors, written as
\begin{eqnarray}
1 &=&\int D\overline{\phi }D\phi D\overline{\omega }D\omega \,\exp \left[
-\int d^{4}\!x\,\left( \overline{\phi }_{\mu }^{ac}\mathcal{M}^{ab}\phi
_{\mu }^{bc}-\overline{\omega }_{\mu }^{ac}\mathcal{M}^{ab}\omega _{\mu
}^{bc}\right) \right] \;,  \nonumber \\
1 &=&\int D\overline{B}DBD\overline{G}DG\,\exp \left[ -\frac{1}{4}\int
d^{4}\!x\,\left( \overline{B}_{\mu \nu }^{a}D_{\sigma }^{ac}D_{\sigma
}^{cb}B_{\mu \nu }^{b}-\overline{G}_{\mu \nu }^{a}D_{\sigma }^{ac}D_{\sigma
}^{cb}G_{\mu \nu }^{b}\right) \right] \;.
\end{eqnarray}
Nevertheless, the action \eqref{local-zero} may be written in a BRST
invariant fashion. To see this, let us first introduce the following
nilpotent BRST transformation
\begin{eqnarray}
sA_{\mu }^{a} &=&-D_{\mu }^{ab}c^{b}\;,  \nonumber \\
sc^{a} &=&\frac{g}{2}f^{abc}c^{b}c^{c}\;,  \nonumber \\
sB_{\mu \nu }^{a} &=&gf^{abc}c^{b}B_{\mu \nu }^{c}+G_{\mu \nu }^{a}\;,
\nonumber \\
sG_{\mu \nu }^{a} &=&gf^{abc}c^{b}G_{\mu \nu }^{c}\;,  \nonumber \\
s\overline{G}_{\mu \nu }^{a} &=&gf^{abc}c^{b}\overline{G}_{\mu \nu }^{c}+%
\overline{B}_{\mu \nu }^{a}\;,  \nonumber \\
s\overline{B}_{\mu \nu }^{a} &=&gf^{abc}c^{b}\overline{B}_{\mu \nu }^{c}\;,
\nonumber \\
s\overline{c}^{a} &=&b^{a}\;,  \nonumber \\
sb^{a} &=&0\;,  \nonumber \\
s\phi _{\mu }^{ab} &=&\omega _{\mu }^{ab}\;,  \nonumber \\
s\omega _{\mu }^{ab} &=&0\;,  \nonumber \\
s\overline{\omega }_{\mu }^{ab} &=&\overline{\phi }_{\mu }^{ab}\;,  \nonumber
\\
s\overline{\phi }_{\mu }^{ab} &=&0\;,  \nonumber  \label{brst} \\
s^{2} &=&0\;.
\end{eqnarray}
Now, let $S_{0}$ be the action defined by
\begin{equation}
S_{0}=S_{\mathrm{YM}}+s\int d^{4}\!x\,(\overline{c}^{a}\partial _{\mu
}A_{\mu }^{a}+\overline{\omega }_{\mu }^{ac}\mathcal{M}^{ab}\phi _{\mu
}^{bc}+\overline{G}_{\mu \nu }^{a}D_{\sigma }^{ac}D_{\sigma }^{cb}B_{\mu \nu
}^{b})\;,  \label{S-zero}
\end{equation}
which satifies
\begin{equation}
sS_{0}=0\;.
\end{equation}
Applying the BRST transformations \eqref{brst} and recalling that the
Faddeev-Popov operator, $\mathcal{M}^{ab}$, is given by \eqref{FPop}, we
obtain
\begin{equation}
S_{0}=S_{\mathrm{Local}}^{\gamma =m=0}+\int d^{4}\!x\,\overline{\omega }%
_{\mu }^{ac}\partial _{\nu }\Bigl(gf^{abd}\phi _{\mu }^{bc}\,D_{\nu
}^{de}c^{e}\Bigr)\;.
\end{equation}
Following \cite{Zwanziger:1992qr} one can show that $S_{0}$ and $S_{\mathrm{%
Local}}^{\gamma =m=0}$ are equivalent. More precisely, one may transform $S_{%
\mathrm{Local}}^{\gamma =m=0}$ into $S_{0}$ by performing the following
shift in the variable $\omega _{\mu }^{ac}$,
\begin{equation}
\omega _{\mu }^{ac}\to \omega _{\mu }^{ac}-\left( \mathcal{M}^{-1}\right)
^{ab}\partial _{\nu }\Bigl(gf^{bed}\phi _{\mu }^{ec}\,D_{\nu }^{dn}c^{n}\Bigr%
)\;,
\end{equation}
and keeping in mind that the corresponding Jacobian turns out to be field
independent. Thus, the following equivalence holds,
\begin{equation}
\int D\Phi \,e^{-S_{0}}=\int D\Phi \,e^{-S_{\mathrm{Local}}^{\gamma =m=0}}\;,
\end{equation}
where $\Phi $ is a shorthand for all the fields. Now, let us reintroduce the
term $S_{\gamma }$, given by \eqref{Sgamma}, while $S_{m}$ remains absent.
It is easy to show, using the BRST transformations \eqref{brst}, that $%
S_{\gamma }$ may be rewritten as
\begin{equation}
S_{\gamma }=\gamma ^{2}\int d^{4}\!x\,\Bigr[gf^{abc}\phi _{\mu }^{ac}A_{\mu
}^{b}-s(gf^{abc}\overline{\omega }_{\mu }^{ac}A_{\mu }^{b})+gf^{abc}%
\overline{\omega }_{\mu }^{ac}\,D_{\mu }^{bd}c^{d}\Bigr]\;.
\end{equation}
The last term can be eliminated by means of a change of variables
\begin{equation}
\omega _{\mu }^{bc}\to \omega _{\mu }^{bc}+\gamma ^{2}\left( \mathcal{M}%
^{-1}\right) ^{bd}gf^{dec}D_{\mu }^{en}c^{n}\;.
\end{equation}
Furthermore, we notice that, thanks to fact that the integral of a total
derivative vanishes, the following expression for $S_{\gamma }$ holds
\begin{equation}
S_{\gamma }=-\gamma ^{2}\int d^{4}\!x\,\Bigr[D_{\mu }^{ab}\phi _{\mu
}^{ba}-s(D_{\mu }^{ab}\overline{\omega }_{\mu }^{ba})\Bigr]\;.
\end{equation}
Nevertheless, the action,
\begin{equation}
S_{0}+S_{\gamma }\;,  \label{not_invariant}
\end{equation}
is not yet BRST invariant. This point can be dealt with by means of the
introduction of a pair of BRST doublets of local external sources \cite
{Zwanziger:1992qr}, $(M_{\mu \nu }^{ab},N_{\mu \nu }^{ab})$ and $(\overline{M%
}_{\mu \nu }^{ab},\overline{N}_{\mu \nu }^{ab})$, which transform as
\begin{eqnarray}
sM_{\mu \nu }^{ab} &=&-N_{\mu \nu }^{ab}\;,\qquad sN_{\mu \nu }^{ab}=0\;,
\nonumber \\
s\overline{N}_{\mu \nu }^{ab} &=&-\overline{M}_{\mu \nu }^{ab}\;,\qquad s%
\overline{M}_{\mu \nu }^{ab}=0\;.
\end{eqnarray}
As pointed out in \cite{Zwanziger:1992qr}, the introduction of these
external sources allows us to promote expression \eqref{not_invariant} to a
BRST invariant action. In fact, let $S_{\mathrm{sources}}$ be the action
\begin{eqnarray}
S_{\mathrm{sources}} &=&s\int d^{4}\!x\,(\overline{N}_{\mu \nu }^{ac}D_{\mu
}^{ab}\phi _{\nu }^{bc}-M_{\mu \nu }^{ac}D_{\mu }^{ab}\overline{\omega }%
_{\nu }^{bc})  \nonumber \\
&=&\int d^{4}\!x\,[-\overline{M}_{\mu \nu }^{ac}D_{\mu }^{ab}\phi _{\nu
}^{bc}-\overline{N}_{\mu \nu }^{ac}s(D_{\mu }^{ab}\phi _{\nu }^{bc})+N_{\mu
\nu }^{ac}D_{\mu }^{ab}\overline{\omega }_{\nu }^{bc}-M_{\mu \nu
}^{ac}s(D_{\mu }^{ab}\overline{\omega }_{\nu }^{bc})]\;,
\end{eqnarray}
which obviously satisfies
\begin{equation}
sS_{\mathrm{sources}}=0\;.
\end{equation}
When the sources $(M_{\mu \nu }^{ab},\overline{M}_{\mu \nu }^{ab},N_{\mu \nu
}^{ab},\overline{N}_{\mu \nu }^{ab})$ attain their physical values \cite
{Zwanziger:1992qr}, defined by
\begin{eqnarray}
\overline{M}_{\mu \nu }^{ab}\Bigl|_{\mathrm{phys}} &=&-M_{\mu \nu }^{ab}\Bigl%
|_{\mathrm{phys}}=-\gamma ^{2}\delta ^{ab}\delta _{\mu \nu }\;,  \nonumber
\label{phys1} \\
N_{\mu \nu }^{ab}\Bigl|_{\mathrm{phys}} &=&\overline{N}_{\mu \nu }^{ab}\Bigl%
|_{\mathrm{phys}}=0\;,
\end{eqnarray}
it immediately follows that
\begin{equation}
S_{\mathrm{sources}}\Bigl|_{\mathrm{phys}}=S_{\gamma }=-\gamma ^{2}\int
d^{4}\!x\,\Bigr[D_{\mu }^{ab}\phi _{\mu }^{ba}-s(D_{\mu }^{ab}\overline{%
\omega }_{\mu }^{ba})\Bigr]\;.
\end{equation}
One sees thus that the use of the external sources $(M_{\mu \nu }^{ab},%
\overline{M}_{\mu \nu }^{ab},N_{\mu \nu }^{ab},\overline{N}_{\mu \nu }^{ab})$
enables us to introduce an extended action $\Sigma _{0}$, given by
\begin{equation}
\Sigma _{0}=S_{0}+S_{\mathrm{sources}}\;,
\end{equation}
which enjoys the important property of being BRST invariant,
\begin{equation}
s\Sigma _{0}=0\;,
\end{equation}
while reducing to expression \eqref{not_invariant} when the sources attain
their physical values, given by \eqref{phys1}. Recapitulating, we have
rewritten
\begin{equation}
\int DADcD\overline{c}Dbe^{-S+\gamma ^{4}H}=\int DADcD\overline{c}DbD%
\overline{\phi }D\phi \overline{\omega }D\omega De^{-S_{0}-S_{\gamma }}\;.
\label{lokaal}
\end{equation}
It is then easily shown, upon combination of \eqref{m3},\eqref{m4}, %
\eqref{not_invariant} and \eqref{lokaal} that the horizon condition is
implemented by requiring that
\begin{equation}
\frac{\partial \Gamma _{\mathrm{GZ}}}{\partial \gamma ^{2}}=0\,,\qquad
\mathrm{with\;}\gamma ^{2}\neq 0\;,  \label{hor2}
\end{equation}
whereby $\Gamma _{\mathrm{GZ}}$ is the Gribov-Zwanziger effective action
defined by
\begin{equation}
e^{-\Gamma _{\mathrm{GZ}}}=\int DADcD\overline{c}DbD\overline{\phi }D\phi D%
\overline{\omega }D\omega e^{-S_{0}-S_{\gamma }}\;.  \label{hor3}
\end{equation}
\noindent To continue, let us analyze the term $S_{m}$, given by \eqref{Sm}.
This term is, just as $S_{\mathcal{O}}^{\mathrm{Local}}$ in \eqref{Llocal-O}%
, left invariant by the gauge transformations \cite{Capri:2005dy}
\begin{eqnarray}
\delta A_{\mu }^{a} &=&-D_{\mu }^{ab}\theta ^{b}\;,  \nonumber  \label{gtm2}
\\
\delta B_{\mu \nu }^{a} &=&gf^{abc}\theta ^{b}B_{\mu \nu }^{c}\;,  \nonumber
\\
\delta G_{\mu \nu }^{a} &=&gf^{abc}\theta ^{b}G_{\mu \nu }^{c}\;,  \nonumber
\\
\delta \overline{G}_{\mu \nu }^{a} &=&gf^{abc}\theta ^{b}\overline{G}_{\mu
\nu }^{c}\;,  \nonumber \\
\delta \overline{B}_{\mu \nu }^{a} &=&gf^{abc}\theta ^{b}\overline{B}_{\mu
\nu }^{c}\;,
\end{eqnarray}
where $\theta ^{a}$ is the parameter of the gauge
transformation, but it is not invariant by the BRST transformations %
\eqref{brst}. This problem can be solved in a way equivalent as done
in the case of $S_{\gamma }$. This time we will introduce a pair of
BRST doublets of external sources, $(U_{\alpha \beta \mu \nu
},V_{\alpha
\beta \mu \nu })$ and $(\overline{U}_{\alpha \beta \mu \nu },\overline{V}%
_{\alpha \beta \mu \nu })$, transforming as
\begin{eqnarray}
sV_{\alpha \beta \mu \nu } &=&U_{\alpha \beta \mu \nu }\;,\qquad sU_{\alpha
\beta \mu \nu }=0\;,  \nonumber \\
s\overline{U}_{\alpha \beta \mu \nu } &=&\overline{V}_{\alpha \beta \mu \nu
}\;,\qquad s\overline{V}_{\alpha \beta \mu \nu }=0\;.
\end{eqnarray}
Hence, by considering the following term
\[
S_{\mathrm{sources}}^{\prime }=s\int d^{4}\!x\,(V_{\alpha \beta \mu \nu }%
\overline{G}_{\alpha \beta }^{a}-\overline{U}_{\alpha \beta \mu \nu
}B_{\alpha \beta }^{a})F_{\mu \nu }^{a}\;,
\]
the term $S_{\mathrm{sources}}^{\prime }$ reduces to $S_{m}$ of \eqref{Sm}
when the sources $(U_{\alpha \beta \mu \nu },\overline{U}_{\alpha \beta \mu
\nu },V_{\alpha \beta \mu \nu },\overline{V}_{\alpha \beta \mu \nu })$
attain the subsequent physical values
\begin{eqnarray}
\overline{V}_{\alpha \beta \mu \nu }\Bigl|_{\mathrm{phys}} &=&V_{\alpha
\beta \mu \nu }\Bigl|_{\mathrm{phys}}=-\frac{im}{2}(\delta _{\alpha \mu
}\delta _{\beta \nu }-\delta _{\alpha \nu }\delta _{\beta \mu })\;,
\nonumber  \label{phys2} \\
U_{\alpha \beta \mu \nu }\Bigl|_{\mathrm{phys}} &=&\overline{U}_{\alpha
\beta \mu \nu }\Bigl|_{\mathrm{phys}}=0\;.
\end{eqnarray}
These sources enable us to define an action $\Sigma _{1}$ as
\begin{equation}
\Sigma _{1}=\Sigma _{0}+S_{\mathrm{sources}}^{\prime }\;,
\end{equation}
in such way that
\begin{equation}
s\Sigma _{1}=0\;.
\end{equation}

\subsection{The global $U(f)$ and $U(f^{\prime})$ symmetries}

In addition to the BRST invariance the action $\Sigma _{1}$ displays global
symmetries $U(f)$, $f=4(N^{2}-1)$ and $U(f^{\prime })$, $f^{\prime }=6$,
respectively expressed by
\begin{eqnarray}
\mathcal{Q}_{\mu \nu }^{ab}(\Sigma _{1}) &\equiv &\int d^{4}\!x\,\biggl( %
\phi _{\mu }^{ca}\frac{\delta \Sigma _{1}}{\delta \phi _{\nu }^{cb}}-%
\overline{\phi }_{\nu }^{cb}\frac{\delta \Sigma _{1}}{\delta \overline{\phi }%
_{\mu }^{ca}}+\omega _{\mu }^{ca}\frac{\delta \Sigma _{1}}{\delta \omega
_{\nu }^{cb}}-\overline{\omega }_{\nu }^{cb}\frac{\delta \Sigma _{1}}{\delta
\overline{\omega }_{\nu }^{ca}}+M_{\sigma \mu }^{ca}\frac{\delta \Sigma _{1}%
}{\delta M_{\sigma \nu }^{cb}}  \nonumber \\
&&-\overline{M}_{\sigma \nu }^{cb}\frac{\delta \Sigma _{1}}{\delta \overline{%
M}_{\sigma \mu }^{ca}}+N_{\sigma \mu }^{ca}\frac{\delta \Sigma _{1}}{\delta
N_{\sigma \nu }^{cb}}-\overline{N}_{\sigma \nu }^{cb}\frac{\delta \Sigma _{1}%
}{\delta \overline{N}_{\sigma \mu }^{ca}}\biggr)=0\;,
\end{eqnarray}
and
\begin{eqnarray}
\mathcal{Q}_{\alpha \beta \mu \nu }(\Sigma _{1}) &\equiv &\int d^{4}\!x\,%
\biggl( B_{\alpha \beta }^{a}\frac{\delta \Sigma _{1}}{\delta B_{\mu \nu
}^{a}}-\overline{B}_{\mu \nu }^{a}\frac{\delta \Sigma _{1}}{\delta \overline{%
B}_{\alpha \beta }^{a}}+G_{\alpha \beta }^{a}\frac{\delta \Sigma _{1}}{%
\delta G_{\mu \nu }^{a}}-\overline{G}_{\mu \nu }^{a}\frac{\delta \Sigma _{1}%
}{\delta \overline{G}_{\alpha \beta }^{a}}+U_{\alpha \beta \sigma \rho }%
\frac{\delta \Sigma _{1}}{\delta U_{\mu \nu \sigma \rho }}  \nonumber \\
&&-\overline{U}_{\mu \nu \sigma \rho }\frac{\delta \Sigma _{1}}{\delta
\overline{U}_{\alpha \beta \sigma \rho }}+V_{\alpha \beta \sigma \rho }\frac{%
\delta \Sigma _{1}}{\delta V_{\mu \nu \sigma \rho }}-\overline{V}_{\mu \nu
\sigma \rho }\frac{\delta \Sigma _{1}}{\delta \overline{V}_{\alpha \beta
\sigma \rho }}\biggr)=0\;,
\end{eqnarray}
The presence of the global invariances $U(f)$ and $U(f^{\prime })$
means that one can make use \cite{Zwanziger:1992qr,Capri:2005dy} of
the composite indices $I\equiv (a,\mu )$, $I=1,\dots ,f$, and
$i\equiv (\mu ,\nu )$, $i=1,\dots ,f^{\prime }$. Specifically,
setting
\begin{eqnarray}
(\phi _{I}^{a},\overline{\phi }_{I}^{a},\omega _{I}^{a},\overline{\omega }%
_{I}^{a}) &\equiv &(\phi _{\mu }^{ab},\overline{\phi }_{\mu }^{ab},\omega
_{\mu }^{ab},\overline{\omega }_{\mu }^{ab})\;,  \nonumber \\
(M_{\mu I}^{a},\overline{M}_{\mu I}^{a},N_{\mu I}^{a},\overline{N}_{\mu
I}^{a}) &\equiv &(M_{\mu \nu }^{ab},\overline{M}_{\mu \nu }^{ab},N_{\mu \nu
}^{ab},\overline{N}_{\mu \nu }^{ab})\;,
\end{eqnarray}
and
\begin{eqnarray}
(B_{i}^{a},\overline{B}_{i}^{a},G_{i}^{a},\overline{G}_{i}^{a}) &\equiv &%
\frac{1}{2}(B_{\mu \nu }^{a},\overline{B}_{\mu \nu }^{a},G_{\mu \nu }^{a},%
\overline{G}_{\mu \nu }^{a})\;,  \nonumber \\
(U_{i\mu \nu },\overline{U}_{i\mu \nu },V_{i\mu \nu },\overline{V}_{i\mu \nu
}) &\equiv &\frac{1}{2}(U_{\alpha \beta \mu \nu },\overline{U}_{\alpha \beta
\mu \nu },V_{\alpha \beta \mu \nu },\overline{V}_{\alpha \beta \mu \nu })\;,
\end{eqnarray}
we rewrite $\Sigma _{1}$ as
\begin{eqnarray}
\Sigma _{1} &=&S_{\mathrm{YM}}+\int d^{4}\!x\,\biggl\{b^{a}\,\partial _{\mu
}A_{\mu }^{a}+\overline{c}^{a}\partial _{\mu }D_{\mu }^{ab}c^{b}+\overline{%
\phi }_{I}^{a}\mathcal{M}^{ab}\phi _{I}^{b}-\overline{\omega }_{I}^{a}%
\mathcal{M}^{ab}\omega _{I}^{b}+gf^{abc}\overline{\omega }_{I}^{a}\partial
_{\mu }(\phi _{I}^{b}D_{\mu }^{cd}c^{d})  \nonumber \\
&&-\overline{M}_{\mu I}^{a}\,D_{\mu }^{ab}\phi _{I}^{b}-\overline{N}_{\mu
I}^{a}\Bigl[D_{\mu }^{ab}\omega _{I}^{b}+gf^{abc}\phi _{I}^{b}D_{\mu
}^{cd}c^{d}\Bigr]+N_{\mu I}^{a}\,D_{\mu }^{ab}\overline{\omega }_{I}^{b}
\nonumber \\
&&-M_{\mu I}^{a}\Bigl[D_{\mu }^{ab}\overline{\phi }_{I}^{b}-gf^{abc}%
\overline{\omega }_{I}^{b}D_{\mu }^{cd}c^{d}\Bigr]+\overline{B}%
_{i}^{a}D_{\mu }^{ab}D_{\mu }^{bc}B_{i}^{c}-\overline{G}_{i}^{a}D_{\mu
}^{ab}D_{\mu }^{bc}G_{i}^{c}  \nonumber \\
&&+F_{\mu \nu }^{a}\bigl(\overline{U}_{i\mu \nu }G_{i}^{a}+V_{i\mu \nu }%
\overline{B}_{i}^{a}-\overline{V}_{i\mu \nu }B_{i}^{a}+U_{i\mu \nu }%
\overline{G}_{i}^{a}\bigr) \biggr\}\;,
\end{eqnarray}
For the symmetry generators, we have
\begin{eqnarray}
\mathcal{Q}_{IJ} &\equiv &\int d^{4}\!x\,\biggl( \phi _{I}^{a}\frac{\delta }{%
\delta \phi _{J}^{a}}-\overline{\phi }_{J}^{a}\frac{\delta }{\delta
\overline{\phi }_{I}^{a}}+\omega _{I}^{a}\frac{\delta }{\delta \omega
_{J}^{a}}-\overline{\omega }_{J}^{a}\frac{\delta }{\delta \overline{\omega }%
_{I}^{a}}+M_{\mu I}^{a}\frac{\delta }{\delta M_{J}^{a}}  \nonumber \\
&&-\overline{M}_{J}^{a}\frac{\delta }{\delta \overline{M}_{I}^{a}}+N_{\mu
I}^{a}\frac{\delta }{\delta N_{\mu J}^{a}}-\overline{N}_{\mu J}^{a}\frac{%
\delta }{\delta \overline{N}_{\mu I}^{a}}\biggr)\;,
\end{eqnarray}
and
\begin{eqnarray}
\mathcal{Q}_{ij} &\equiv &\int d^{4}\!x\,\biggl( B_{i}^{a}\frac{\delta }{%
\delta B_{j}^{a}}-\overline{B}_{j}^{a}\frac{\delta }{\delta \overline{B}%
_{i}^{a}}+G_{i}^{a}\frac{\delta }{\delta G_{j}^{a}}-\overline{G}_{j}^{a}%
\frac{\delta }{\delta \overline{G}_{i}^{a}}+U_{i\mu \nu }\frac{\delta }{%
\delta U_{j\mu \nu }}-\overline{U}_{j\mu \nu }\frac{\delta }{\delta
\overline{U}_{i\mu \nu }}  \nonumber \\
&&+V_{i\mu \nu }\frac{\delta }{\delta V_{j\mu \nu }}-\overline{V}_{j\mu \nu }%
\frac{\delta }{\delta \overline{V}_{i\mu \nu }}\biggr)\;.
\end{eqnarray}
By means of the trace of these operators the $I(i)$-valued fields turn out
to possess an additional quantum number, displayed in Tables \ref{table1}
and \ref{table2}, together with the dimension and the ghost number.
\begin{table}[t]
\centering
\begin{tabular}{|l|cccccccccccc|}
\hline
& $A$ & $b$ & $\overline c$ & $c$ & $\phi$ & $\overline\phi$ & $\omega$ & $%
\overline\omega\phantom{\Bigl|}$ & $B$ & $\overline B$ & $G$ & $\overline G$
\\ \hline\hline
dimension & $1$ & $2$ & $2$ & $0$ & $1$ & $1$ & $1$ & $1$ & $1$ & $1$ & $1$
& $1$ \\
ghost number & $0$ & $0$ & $-1$ & $1$ & $0$ & $0$ & $1$ & $-1$ & $0$ & $0$ &
$1$ & $-1$ \\
$Q_f$-charge & $0$ & $0$ & $0$ & $0$ & $1$ & $-1$ & $1$ & $-1$ & $0$ & $0$ &
$0$ & $0$ \\
$Q_{f^{\prime}}$-charge & $0$ & $0$ & $0$ & $0$ & $0$ & $0$ & $0$ & $0$ & $1$
& $-1$ & $1$ & $-1$ \\ \hline
\end{tabular}
\caption{Quantum numbers of the fields}
\label{table1}
\end{table}
\begin{table}[t]
\centering
\begin{tabular}{|l|cccccccc|}
\hline
& $\phantom{\Bigl|}M$ & $\overline M$ & $N$ & $\overline N$ & $V$ & $%
\overline V$ & $U$ & $\overline U$ \\ \hline\hline
dimension & $2$ & $2$ & $2$ & $2$ & $1$ & $1$ & $1$ & $1$ \\
ghost number & $0$ & $0$ & $1$ & $-1$ & $0$ & $0$ & $1$ & $-1$ \\
$Q_f$-charge & $1$ & $-1$ & $1$ & $-1$ & $0$ & $0$ & $0$ & $0$ \\
$Q_{f^{\prime}}$-charge & $0$ & $0$ & $0$ & $0$ & $1$ & $-1$ & $1$ & $-1$ \\
\hline
\end{tabular}
\caption{Quantum numbers of the sources}
\label{table2}
\end{table}

\subsection{The complete classical action $\Sigma$}

We proceed by establishing the complete set of Ward identities which
will enable us to analyze the renormalizability of the theory to all
orders. Let us first identify the final complete action to start
with. For this purpose, we need to introduce additional external
sources $(\Omega^{a}_{\mu}, L^{a}, \overline Y^{a}_{i},
Y^{a}_{i},\overline X^{a}_{i}, X^{a}_{i} )$ in order to define at
quantum level the composite operators entering the nonlinear BRST
transformations of the fields $(A^{a}_{\mu}, L^{a},
B^{a}_{i},\overline B^{a}_{i},G^{a}_{i}, \overline G^{a}_{i})$, eqs
\eqref{brst}. In the present case this term reads
\begin{equation}  \label{ext}
S_{\mathrm{ext}}=s\int d^{4}\!x\,(-\Omega^{a}_{\mu}A^{a}_{\mu}
+L^{a}c^{a}-\overline Y^{a}_{i}B^{a}_{i} -Y^{a}_{i}\overline B^{a}_{i}
+\overline X^{a}_{i}G^{a}_{i}+X^{a}_{i}\overline G^{a}_{i})\;,
\end{equation}
with
\begin{equation}
s\Omega^{a}_{\mu}=sL^{a}=0\;,
\end{equation}
and
\begin{eqnarray}
sY^{a}_{i}&=&X^{a}_{i}\;,  \nonumber \\
sX^{a}_{i}&=&0\;,  \nonumber \\
s\overline X^{a}_{i}&=&-\overline Y^{a}_{i}\;,  \nonumber \\
s\overline Y^{a}_{i}&=&0\;.
\end{eqnarray}
The quantum numbers of the external sources $(\Omega^{a}_{\mu}, L^{a},
\overline Y^{a}_{i}, Y^{a}_{i},\overline X^{a}_{i}, X^{a}_{i} )$ are
displayed in Table \ref{table3}.
\begin{table}[t]
\centering
\begin{tabular}{|l|cccccc|}
\hline
& $X$ & $\overline X$ & $Y$ & $\overline Y$ & $\Omega$ & $L\phantom{\Bigl|}$
\\ \hline\hline
dimension & $3$ & $3$ & $3$ & $3$ & $3$ & $4$ \\
ghost number & $0$ & $-2$ & $-1$ & $-1$ & $-1$ & $-2$ \\
$Q_f$-charge & $0$ & $0$ & $0$ & $0$ & $0$ & $0$ \\
$Q_{f^{\prime}}$-charge & $1$ & $-1$ & $1$ & $-1$ & $0$ & $0$ \\ \hline
\end{tabular}
\caption{Quantum numbers of the external sources}
\label{table3}
\end{table}

\noindent Furthermore, we have to add the extra source term $S_{\mathrm{extra%
}}$ for renormalization purposes, as it was explained in \cite
{Zwanziger:1992qr,Capri:2005dy}
\begin{eqnarray}
S_{\mathrm{extra}}&=&\int d^{4}\!x\,\biggl\{\overline M^{a}_{\mu
I}M^{a}_{\mu I} -\overline N^{a}_{\mu I}N^{a}_{\mu I} +\lambda_{1}(\overline
B^{a}_{i}B^{a}_{i}-\overline G^{a}_{i}G^{a}_{i}) (\overline
V_{j\mu\nu}V_{j\mu\nu}-\overline U_{j\mu\nu}U_{j\mu\nu})  \nonumber \\
&&+\frac{\lambda^{abcd}}{16}(\overline B^{a}_{i}B^{b}_{i}-\overline
G^{a}_{i}G^{b}_{i})(\overline B^{c}_{j}B^{d}_{j}-\overline
G^{c}_{j}G^{d}_{j}) +\lambda_{3}\Bigl(\overline
B^{a}_{i}G^{a}_{j}V_{i\mu\nu}\overline U_{j\mu\nu} +\overline
G^{a}_{i}G^{a}_{j}U_{i\mu\nu}\overline U_{j\mu\nu}  \nonumber \\
&&+\overline B^{a}_{i}B^{a}_{j}V_{i\mu\nu}\overline V_{j\mu\nu} -\overline
G^{a}_{i}B^{a}_{j}V_{j\mu\nu}U_{i\mu\nu} -G^{a}_{i}B^{a}_{j}\overline
U_{i\mu\nu}\overline V_{j\mu\nu} +\overline G^{a}_{i}\overline
B^{a}_{j}U_{i\mu\nu}V_{j\mu\nu}  \nonumber \\
&&-\frac{1}{2}B^{a}_{i}B^{a}_{j}\overline V_{i\mu\nu}\overline V_{j\mu\nu} +%
\frac{1}{2}G^{a}_{i}G^{a}_{j}\overline U_{i\mu\nu}\overline U_{j\mu\nu} -%
\frac{1}{2}\overline B^{a}_{i}\overline B^{a}_{j}V_{i\mu\nu}V_{j\mu\nu} +%
\frac{1}{2}\overline G^{a}_{i}\overline G^{a}_{j}U_{i\mu\nu}U_{j\mu\nu}\Bigr)
\nonumber \\
&& +\chi_{1}(\overline V_{i\mu\nu}\partial^{2}V_{i\mu\nu} -\overline
U_{i\mu\nu}\partial^{2}U_{i\mu\nu}) +\chi_{2}(\overline
V_{i\mu\nu}\partial_{\mu}\partial_{\alpha}V_{i\nu\alpha} -\overline
U_{i\mu\nu}\partial_{\mu}\partial_{\alpha}U_{i\nu\alpha})  \nonumber \\
&&-\zeta(\overline U_{i\mu\nu}U_{i\mu\nu}\overline
U_{j\alpha\beta}U_{j\alpha\beta} +\overline V_{i\mu\nu}V_{i\mu\nu}\overline
V_{j\alpha\beta}V _{j\alpha\beta} -2\overline
U_{i\mu\nu}U_{i\mu\nu}\overline V_{j\alpha\beta}V_{j\alpha\beta}) \biggr\}\;,
\end{eqnarray}
where $\lambda_{1},\lambda_{3},\chi_{1},\chi_{2},\zeta$ are free parameters,
and the gauge invariant rank 4 tensor $\lambda^{abcd}$ has the following
symmetry properties
\begin{equation}
\lambda^{abcd}=\lambda^{cdab}=\lambda^{bacd}\;,
\end{equation}
and it obeys a generalized Jacobi identity
\begin{equation}
f^{man}\lambda^{mbcd}+f^{mbn}\lambda^{amcd}
+f^{mcn}\lambda^{abmd}+f^{mdn}\lambda^{abcn}=0\;.
\end{equation}
Thus, the complete action we are looking for is
\begin{eqnarray}
\Sigma &=&\Sigma_{1}+S_{\mathrm{ext}}+S_{\mathrm{extra}}  \nonumber \\
&=&S_{\mathrm{YM}} +\int d^{4}\!x\,\biggl\{b^{a}\,\partial_{\mu}A^{a}_{\mu}
+\overline c^{a}\partial_{\mu}D^{ab}_{\mu}c^{b} +\overline\phi^{a}_{I}%
\mathcal{M}^{ab}\phi^{b}_{I} -\overline\omega^{a}_{I}\mathcal{M}%
^{ab}\omega^{b}_{I}
+gf^{abc}\overline\omega^{a}_{I}\partial_{\mu}(\phi^{b}_{I}D^{cd}_{\mu}c^{d})
\nonumber \\
&&-\overline M^{a}_{\mu I}\,D^{ab}_{\mu}\phi^{b}_{I} -\overline N^{a}_{\mu I}%
\Bigl[D^{ab}_{\mu}\omega^{b}_{I} +gf^{abc}\phi^{b}_{I}D^{cd}_{\mu}c^{d}\Bigr%
] + N^{a}_{\mu I}\,D^{ab}_{\mu}\overline\omega^{b}_{I}  \nonumber \\
&&- M^{a}_{\mu I}\Bigl[D^{ab}_{\mu}\overline\phi^{b}_{I}
-gf^{abc}\overline\omega^{b}_{I}D^{cd}_{\mu}c^{d}\Bigr] +\overline
M^{a}_{\mu I}M^{a}_{\mu I} -\overline N^{a}_{\mu I}N^{a}_{\mu I} +\overline
B^{a}_{i}D^{ab}_{\mu}D^{bc}_{\mu}B^{c}_{i}  \nonumber \\
&&-\overline G^{a}_{i}D^{ab}_{\mu}D^{bc}_{\mu}G^{c}_{i} +F^{a}_{\mu\nu}\bigl(%
\overline U_{i\mu\nu}G^{a}_{i} +V_{i\mu\nu}\overline B^{a}_{i} -\overline
V_{i\mu\nu}B^{a}_{i} +U_{i\mu\nu}\overline G^{a}_{i}\bigr)  \nonumber \\
&& +\lambda_{1}(\overline B^{a}_{i}B^{a}_{i}-\overline G^{a}_{i}G^{a}_{i})
(\overline V_{j\mu\nu}V_{j\mu\nu}-\overline U_{j\mu\nu}U_{j\mu\nu})
\nonumber
\end{eqnarray}
\begin{eqnarray}
&&+\frac{\lambda^{abcd}}{16}(\overline B^{a}_{i}B^{b}_{i}-\overline
G^{a}_{i}G^{b}_{i}) (\overline B^{c}_{j}B^{d}_{j}-\overline
G^{c}_{j}G^{d}_{j}) +\lambda_{3}\Bigl(\overline
B^{a}_{i}G^{a}_{j}V_{i\mu\nu}\overline U_{j\mu\nu} +\overline
G^{a}_{i}G^{a}_{j}U_{i\mu\nu}\overline U_{j\mu\nu}  \nonumber \\
&& +\overline B^{a}_{i}B^{a}_{j}V_{i\mu\nu}\overline V_{j\mu\nu} -\overline
G^{a}_{i}B^{a}_{j}V_{j\mu\nu}U_{i\mu\nu} -G^{a}_{i}B^{a}_{j}\overline
U_{i\mu\nu}\overline V_{j\mu\nu} +\overline G^{a}_{i}\overline
B^{a}_{j}U_{i\mu\nu}V_{j\mu\nu}  \nonumber \\
&&-\frac{1}{2}B^{a}_{i}B^{a}_{j}\overline V_{i\mu\nu}\overline V_{j\mu\nu}+%
\frac{1}{2}G^{a}_{i}G^{a}_{j}\overline U_{i\mu\nu}\overline U_{j\mu\nu} -%
\frac{1}{2}\overline B^{a}_{i}\overline B^{a}_{j}V_{i\mu\nu}V_{j\mu\nu} +%
\frac{1}{2}\overline G^{a}_{i}\overline G^{a}_{j}U_{i\mu\nu}U_{j\mu\nu}\Bigr)
\nonumber \\
&& +\chi_{1}(\overline V_{i\mu\nu}\partial^{2}V_{i\mu\nu} -\overline
U_{i\mu\nu}\partial^{2}U_{i\mu\nu}) +\chi_{2}(\overline
V_{i\mu\nu}\partial_{\mu}\partial_{\alpha}V_{i\nu\alpha} -\overline
U_{i\mu\nu}\partial_{\mu}\partial_{\alpha}U_{i\nu\alpha})  \nonumber \\
&&-\zeta(\overline U_{i\mu\nu}U_{i\mu\nu}\overline
U_{j\alpha\beta}U_{j\alpha\beta} +\overline V_{i\mu\nu}V_{i\mu\nu}\overline
V_{j\alpha\beta}V _{j\alpha\beta} -2\overline
U_{i\mu\nu}U_{i\mu\nu}\overline V_{j\alpha\beta}V_{j\alpha\beta})
-\Omega^{a}_{\mu}D^{ab}_{\mu}c^{b}  \nonumber \\
&&+\frac{g}{2}f^{abc}L^{a}c^{b}c^{c}+gf^{abc}\overline
Y^{a}_{i}c^{b}B^{c}_{i} +gf^{abc}Y^{a}_{i}c^{b}\overline B^{c}_{i}
+gf^{abc}\overline X^{a}_{i}c^{b}G^{c}_{i} +gf^{abc}X^{a}_{i}c^{b}\overline
G^{c}_{i}\biggr\}\;.  \label{sigma}
\end{eqnarray}

\sect{The complete set of Ward identities}

In this section, we have enlisted all known Ward identities, associated to
the action \eqref{sigma}.

\begin{itemize}
\item  {The Slavnov-Taylor identity
\begin{eqnarray}
&&\mathcal{S}(\Sigma )\equiv   \nonumber \\
&&\int d^{4}\!x\,\biggl[\frac{\delta \Sigma }{\delta \Omega _{\mu }^{a}}%
\frac{\delta \Sigma }{\delta A_{\mu }^{a}}+\frac{\delta \Sigma }{\delta L^{a}%
}\frac{\delta \Sigma }{\delta c^{a}}+b^{a}\frac{\delta \Sigma }{\delta
\overline{c}^{a}}+\omega _{I}^{a}\frac{\delta \Sigma }{\delta \phi _{I}^{a}}+%
\overline{\phi }_{I}^{a}\frac{\delta \Sigma }{\delta \overline{\omega }%
_{I}^{a}}-\overline{M}_{\mu I}^{a}\frac{\delta \Sigma }{\delta \overline{N}%
_{\mu I}^{a}}-N_{\mu I}^{a}\frac{\delta \Sigma }{\delta M_{\mu I}^{a}}
\nonumber \\
&&+\biggl(\frac{\delta \Sigma }{\delta \overline{Y}_{i}^{a}}+G_{i}^{a}%
\biggr) \frac{\delta \Sigma }{\delta B_{i}^{a}}+\frac{\delta \Sigma }{\delta
Y_{i}^{a}}\frac{\delta \Sigma }{\delta \overline{B}_{i}^{a}}+\frac{\delta
\Sigma }{\delta \overline{X}_{i}^{a}}\frac{\delta \Sigma }{\delta G_{i}^{a}}+%
\biggl(\frac{\delta \Sigma }{\delta X_{i}^{a}}+\overline{B}_{i}^{a}\biggr)
\frac{\delta \Sigma }{\delta \overline{G}_{i}^{a}}+\overline{V}_{i\mu \nu }%
\frac{\delta \Sigma }{\delta \overline{U}_{i\mu \nu }}  \nonumber \\
&&+U_{i\mu \nu }\frac{\delta \Sigma }{\delta V_{i\mu \nu }}-\overline{Y}%
_{i}^{a}\frac{\delta \Sigma }{\delta \overline{X}_{i}^{a}}+X_{i}^{a}\frac{%
\delta \Sigma }{\delta Y_{i}^{a}}\biggr]=0\;,  \label{slavnov}
\end{eqnarray}
}

\item  {The Landau gauge fixing
\begin{equation}
\frac{\delta \Sigma }{\delta b^{a}}=\partial _{\mu }A_{\mu }^{a}\;.
\end{equation}
}

\item  {The antighost equation
\begin{equation}
\frac{\delta \Sigma }{\delta \overline{c}^{a}}+\partial _{\mu }\frac{\delta
\Sigma }{\delta \Omega _{\mu }^{a}}=0\;.
\end{equation}
}

\item  {The ghost equation
\begin{eqnarray}
\mathcal{G}^{a}(\Sigma ) &\equiv &\int d^{4}\!x\,\biggl[ \frac{\delta \Sigma
}{\delta c^{a}}+gf^{abc}\biggl(\overline{c}^{b}\frac{\delta \Sigma }{\delta
b^{c}}+\phi _{I}^{b}\frac{\delta \Sigma }{\delta \omega _{I}^{c}}+\overline{%
\omega }_{I}^{b}\frac{\delta \Sigma }{\delta \overline{\phi }_{I}^{c}}-%
\overline{N}_{\mu I}^{b}\frac{\delta \Sigma }{\delta \overline{M}_{\mu I}^{c}%
}-M_{\mu I}^{b}\frac{\delta \Sigma }{\delta N_{\mu I}^{c}}\biggl)\biggr]
\nonumber \\
&=&\Delta _{\mathrm{class}}^{a}  \nonumber \\
&=&\int d^{4}\!x\,gf^{abc}\Bigl(\Omega _{\mu }^{b}A_{\mu }^{c}-L^{b}c^{c}+%
\overline{Y}_{i}^{b}B_{i}^{c}+Y_{i}^{b}\overline{B}_{i}^{c}-\overline{X}%
_{i}^{b}G_{i}^{c}-X_{i}^{b}\overline{G}_{i}^{c}\Bigr)\;.
\end{eqnarray}
}

\item  {The rigid group transformations
\begin{eqnarray}
\mathcal{W}^{a}(\Sigma ) &\equiv &gf^{abc}\int d^{4}\!x\,\sum_{k}\psi
_{k}^{b}\frac{\delta \Sigma }{\delta \psi _{k}^{c}}=0\;,  \nonumber \\
\psi _{k}^{a} &\equiv &(A,b,c,\overline{c},\phi ,\overline{\phi },\omega ,%
\overline{\omega },B,\overline{B},G,\overline{G},\Omega ,L,  \nonumber \\
&&M,\overline{M},N,\overline{N},X,\overline{X},Y,\overline{Y})\;.
\end{eqnarray}
}

\item  {The $SL(2,\mathbb{R})$ invariance\footnote{%
See also \cite{Dudal:2002ye}.}
\begin{equation}
\mathcal{D}(\Sigma )\equiv \int d^{4}\!x\,\biggl(c^{a}\frac{\delta \Sigma }{%
\delta \overline{c}^{a}}+\frac{\delta \Sigma }{\delta b^{a}}\frac{\delta
\Sigma }{\delta L^{a}}\biggr)=0\;.  \label{sl2r}
\end{equation}
}

\item  {The $\overline{\phi }$-equation
\begin{equation}
\frac{\delta \Sigma }{\delta \overline{\phi }_{I}^{a}}-\partial _{\mu }\frac{%
\delta \Sigma }{\delta \overline{M}_{\mu I}^{a}}=(1+\chi )\partial _{\mu
}M_{\mu I}^{a}-gf^{abc}M_{\mu I}^{b}A_{\mu }^{c}\;.
\end{equation}
}

\item  {The $\omega $-equation
\begin{equation}
\frac{\delta \Sigma }{\delta \omega _{I}^{a}}+\partial _{\mu }\frac{\delta
\Sigma }{\delta N_{\mu I}^{a}}+gf^{abc}\overline{\omega }_{I}^{b}\frac{%
\delta \Sigma }{\delta b^{c}}=-(1+\chi )\partial _{\mu }\overline{N}_{\mu
I}^{a}+gf^{abc}\overline{N}_{\mu I}^{b}A_{\mu }^{c}\;.
\end{equation}
}

\item  {The $\phi $-equation
\begin{equation}
\frac{\delta \Sigma }{\delta \phi _{I}^{a}}-\partial _{\mu }\frac{\delta
\Sigma }{\delta M_{\mu I}^{a}}+gf^{abc}\biggl(\overline{\phi }_{I}^{b}\frac{%
\delta \Sigma }{\delta b^{c}}+\overline{\omega }_{I}^{b}\frac{\delta \Sigma
}{\delta \overline{c}^{c}}+\overline{N}_{\mu I}^{b}\frac{\delta \Sigma }{%
\delta \Omega _{\mu }^{c}}\biggr) =(1+\chi )\partial _{\mu }\overline{M}%
_{\mu I}^{a}-gf^{abc}\overline{M}_{\mu I}^{b}A_{\mu }^{c}\;.
\end{equation}
}

\item  {The $\overline{\omega }$-equation
\begin{equation}
\frac{\delta \Sigma }{\delta \overline{\omega }_{I}^{a}}+\partial _{\mu }%
\frac{\delta \Sigma }{\delta \overline{N}_{\mu I}^{a}}-gf^{abc}M_{\mu I}^{b}%
\frac{\delta \Sigma }{\delta \Omega _{\mu }^{c}}=(1+\chi )\partial _{\mu
}N_{\mu I}^{a}-gf^{abc}N_{\mu I}^{b}A_{\mu }^{c}\;.
\end{equation}
}

\item  {The global $U(f)$ invariance, $f=4(N^{2}-1)$,
\begin{eqnarray}
\mathcal{Q}_{IJ}(\Sigma ) &\equiv &\int d^{4}\!x\,\biggl( \phi _{I}^{a}\frac{%
\delta \Sigma }{\delta \phi _{J}^{a}}-\overline{\phi }_{J}^{a}\frac{\delta
\Sigma }{\delta \overline{\phi }_{I}^{a}}+\omega _{I}^{a}\frac{\delta \Sigma
}{\delta \omega _{J}^{a}}-\overline{\omega }_{J}^{a}\frac{\delta \Sigma }{%
\delta \overline{\omega }_{I}^{a}}+M_{\mu I}^{a}\frac{\delta \Sigma }{\delta
M_{J}^{a}}  \nonumber \\
&&-\overline{M}_{J}^{a}\frac{\delta \Sigma }{\delta \overline{M}_{I}^{a}}%
+N_{\mu I}^{a}\frac{\delta \Sigma }{\delta N_{\mu J}^{a}}-\overline{N}_{\mu
J}^{a}\frac{\delta \Sigma }{\delta \overline{N}_{\mu I}^{a}}\biggr)=0\;.
\end{eqnarray}
}

\item  {The rigid symmetry related to the horizon function
\[
\mathcal{R}_{IJ}(\Sigma )\equiv \int d^{4}\!x\,\biggl( \phi _{I}^{a}\frac{%
\delta \Sigma }{\delta \omega _{J}^{a}}-\overline{\omega }_{J}^{a}\frac{%
\delta \Sigma }{\delta \overline{\phi }_{I}^{a}}-M_{\mu I}^{a}\frac{\delta
\Sigma }{\delta N_{\mu J}^{a}}+\overline{N}_{\mu J}^{a}\frac{\delta \Sigma }{%
\delta \overline{M}_{\mu I}^{a}}\biggr)=0\;.
\]
}

\item  {The symmetries relating the auxiliary fields $\phi ,\overline{\phi }%
,\omega ,\overline{\omega }$ to the Faddeev-Popov ghost and antighost $c,%
\overline{c}$
\begin{eqnarray}
\mathcal{W}_{I}(\Sigma ) &\equiv &\int d^{4}\!x\,\biggl( \overline{\omega }%
_{I}^{a}\frac{\delta \Sigma }{\delta \overline{c}^{a}}-c^{a}\frac{\delta
\Sigma }{\delta \omega _{I}^{a}}+\overline{N}_{\mu I}^{a}\frac{\delta \Sigma
}{\delta \Omega _{\mu }^{a}}\biggr)=0\;,  \label{W} \\
\mathcal{Q}_{I}(\Sigma ) &\equiv &\int d^{4}\!x\,\biggl( \overline{\phi }%
_{I}^{a}\frac{\delta \Sigma }{\delta \overline{c}^{a}}+c^{a}\frac{\delta
\Sigma }{\delta \phi _{I}^{a}}-M_{\mu I}^{a}\frac{\delta \Sigma }{\delta
\Omega _{\mu }^{a}}+\frac{\delta \Sigma }{\delta L^{a}}\frac{\delta \Sigma }{%
\delta \omega _{I}^{a}}\biggr)=0\;.  \label{Q}
\end{eqnarray}
}

\item  {The global $U(6)$ invariance
\begin{eqnarray}
\mathcal{Q}_{ij}(\Sigma ) &\equiv &\int d^{4}\!x\,\biggl( B_{i}^{a}\frac{%
\delta \Sigma }{\delta B_{j}^{a}}-\overline{B}_{j}^{a}\frac{\delta \Sigma }{%
\delta \overline{B}_{i}^{a}}+G_{i}^{a}\frac{\delta \Sigma }{\delta G_{j}^{a}}%
-\overline{G}_{j}^{a}\frac{\delta \Sigma }{\delta \overline{G}_{i}^{a}}%
+U_{i\mu \nu }\frac{\delta \Sigma }{\delta U_{j\mu \nu }}-\overline{U}_{j\mu
\nu }\frac{\delta \Sigma }{\delta \overline{U}_{i\mu \nu }}  \nonumber \\
&&+V_{i\mu \nu }\frac{\delta \Sigma }{\delta V_{j\mu \nu }}-\overline{V}%
_{j\mu \nu }\frac{\delta \Sigma }{\delta \overline{V}_{i\mu \nu }}+Y_{i}^{a}%
\frac{\delta \Sigma }{\delta Y_{j}^{a}}-\overline{Y}_{j}^{a}\frac{\delta
\Sigma }{\delta \overline{Y}_{i}^{a}}+X_{i}^{a}\frac{\delta \Sigma }{\delta
X_{j}^{a}}-\overline{X}_{j}^{a}\frac{\delta \Sigma }{\delta \overline{X}%
_{i}^{a}}\biggr)=0\;.  \nonumber\\
\end{eqnarray}}
\item  {The rigid symmetries related to the mass operator
\begin{eqnarray}
\mathcal{R}_{ij}^{(\alpha )}(\Sigma ) &=&0\,,\alpha \in \{1,2,3,4\}
\nonumber \\
&&\mbox{with}  \nonumber \\
\mathcal{R}_{ij}^{(1)}(\Sigma ) &\equiv &\int d^{4}\!x\,\biggl( B_{i}^{a}%
\frac{\delta \Sigma }{\delta G_{j}^{a}}-\overline{G}_{j}^{a}\frac{\delta
\Sigma }{\delta \overline{B}_{i}^{a}}+V_{i\mu \nu }\frac{\delta \Sigma }{%
\delta U_{j\mu \nu }}-\overline{U}_{j\mu \nu }\frac{\delta \Sigma }{\delta
\overline{V}_{i\mu \nu }}+Y_{i}^{a}\frac{\delta \Sigma }{\delta X_{j}^{a}}+%
\overline{X}_{j}^{a}\frac{\delta \Sigma }{\delta \overline{Y}_{i}^{a}}\biggr)%
\;,  \nonumber \\
\mathcal{R}_{ij}^{(2)}(\Sigma ) &\equiv &\int d^{4}\!x\,\biggl( \overline{B}%
_{i}^{a}\frac{\delta \Sigma }{\delta \overline{G}_{j}^{a}}+G_{j}^{a}\frac{%
\delta \Sigma }{\delta B_{i}^{a}}+\overline{V}_{i\mu \nu }\frac{\delta
\Sigma }{\delta \overline{U}_{j\mu \nu }}+U_{j\mu \nu }\frac{\delta \Sigma }{%
\delta V_{i\mu \nu }}-\overline{Y}_{i}^{a}\frac{\delta \Sigma }{\delta
\overline{X}_{j}^{a}}-X_{j}^{a}\frac{\delta \Sigma }{\delta Y_{i}^{a}}\biggr)%
\;,  \nonumber \\
\mathcal{R}_{ij}^{(3)}(\Sigma ) &\equiv &\int d^{4}\!x\,\biggl( \overline{B}%
_{i}^{a}\frac{\delta \Sigma }{\delta G_{j}^{a}}-\overline{G}_{j}^{a}\frac{%
\delta \Sigma }{\delta B_{i}^{a}}-\overline{V}_{i\mu \nu }\frac{\delta
\Sigma }{\delta U_{j\mu \nu }}+\overline{U}_{j\mu \nu }\frac{\delta \Sigma }{%
\delta V_{i\mu \nu }}+\overline{Y}_{i}^{a}\frac{\delta \Sigma }{\delta
X_{j}^{a}}+\overline{X}_{j}^{a}\frac{\delta \Sigma }{\delta Y_{i}^{a}}\biggr)%
\;,  \nonumber \\
\mathcal{R}_{ij}^{(4)}(\Sigma ) &\equiv &\int d^{4}\!x\,\biggl( B_{i}^{a}%
\frac{\delta \Sigma }{\delta \overline{G}_{j}^{a}}+G_{j}^{a}\frac{\delta
\Sigma }{\delta \overline{B}_{i}^{a}}-V_{i\mu \nu }\frac{\delta \Sigma }{%
\delta \overline{U}_{j\mu \nu }}-U_{j\mu \nu }\frac{\delta \Sigma }{\delta
\overline{V}_{i\mu \nu }}-Y_{i}^{a}\frac{\delta \Sigma }{\delta \overline{X}%
_{j}^{a}}+X_{j}^{a}\frac{\delta \Sigma }{\delta \overline{Y}_{i}^{a}}\biggr)%
\;.  \nonumber \\
&&
\end{eqnarray}
}
\end{itemize}

\sect{The linearized operators and (anti-)commutation relations} In
order to facilitate the upcoming vast amount of algebra required for
the determination of the most general counterterm, we shall give her
some (anti-)commutation relations between several (linearized)
symmetry operators.

\noindent The equations \eqref{slavnov}, \eqref{sl2r} and \eqref{Q}
generate, respectively, the following linearized operators:
\begin{eqnarray}
\mathcal{B}_{\Sigma}&\equiv&\int d^{4}\!x\,\biggl[\frac{\delta\Sigma}{%
\delta\Omega^{a}_{\mu}} \frac{\delta}{\delta A^{a}_{\mu}} +\frac{\delta\Sigma%
}{\delta A^{a}_{\mu}} \frac{\delta}{\delta\Omega^{a}_{\mu}} +\frac{%
\delta\Sigma}{\delta L^{a}}\frac{\delta}{\delta c^{a}} +\frac{\delta\Sigma}{%
\delta c^{a}}\frac{\delta}{\delta L^{a}} +b^{a}\frac{\delta}{\delta\overline
c^{a}} +\omega^{a}_{I}\frac{\delta}{\delta\phi^{a}_{I}} +\overline%
\phi^{a}_{I}\frac{\delta}{\delta\overline\omega^{a}_{I}}  \nonumber \\
&&-\overline M^{a}_{\mu I}\frac{\delta}{\delta\overline N^{a}_{\mu I}}
-N^{a}_{\mu I}\frac{\delta}{\delta M^{a}_{\mu I}} +\biggl(\frac{\delta\Sigma%
}{\delta\overline Y^{a}_{i}}+G^{a}_{i}\biggr) \frac{\delta}{\delta B^{a}_{i}}
+\frac{\delta\Sigma}{\delta B^{a}_{i}} \frac{\delta}{\delta\overline
Y^{a}_{i}} +\frac{\delta\Sigma}{\delta Y^{a}_{i}} \frac{\delta}{%
\delta\overline B^{a}_{i}} +\frac{\delta\Sigma}{\delta\overline B^{a}_{i}}
\frac{\delta}{\delta Y^{a}_{i}}  \nonumber \\
&& +\frac{\delta\Sigma}{\delta\overline X^{a}_{i}} \frac{\delta}{\delta
G^{a}_{i}} +\frac{\delta\Sigma}{\delta G^{a}_{i}} \frac{\delta}{%
\delta\overline X^{a}_{i}} +\biggl(\frac{\delta\Sigma}{\delta X^{a}_{i}}%
+\overline B^{a}_{i}\biggr) \frac{\delta}{\delta\overline G^{a}_{i}} +\frac{%
\delta\Sigma}{\delta\overline G^{a}_{i}} \frac{\delta}{\delta X^{a}_{i}}
+\overline V_{i\mu\nu}\frac{\delta}{\delta\overline U_{i\mu\nu}} +U_{i\mu\nu}%
\frac{\delta}{\delta V_{i\mu\nu}}  \nonumber \\
&&-\overline Y^{a}_{i}\frac{\delta}{\delta\overline X^{a}_{i}} +X^{a}_{i}%
\frac{\delta}{\delta Y^{a}_{i}}\biggr]\;,  \label{linearized} \\
\cr{\cal D}_{\Sigma}&\equiv&\int d^{4}\!x\,\biggl(c^{a} \frac{\delta}{%
\delta\overline c^{a}}+\frac{\delta\Sigma}{\delta b^{a}} \frac{\delta}{%
\delta L^{a}} +\frac{\delta\Sigma}{\delta L^{a}} \frac{\delta}{\delta b^{a}}%
\biggr)\;, \\
\cr{\cal Q}^{\Sigma}_{I}&\equiv&\int d^{4}\!x\,\biggl( \overline\phi^{a}_{I}%
\frac{\delta}{\delta\overline c^{a}} +c^{a}\frac{\delta}{\delta\phi^{a}_{I}}
-M^{a}_{\mu I}\frac{\delta}{\delta\Omega^{a}_{\mu}} +\frac{\delta\Sigma}{%
\delta L^{a}}\frac{\delta}{\delta\omega^{a}_{I}} +\frac{\delta\Sigma}{%
\delta\omega^{a}_{I}}\frac{\delta}{\delta L^{a}} \biggr)=0\;.
\end{eqnarray}
Consequently, we are able to derive some useful (anti-)commutations
relations:
\begin{eqnarray}
\left[\frac{\delta}{\delta b^{a}}\;,\;\mathcal{B}_{\Sigma}\right]&=& \frac{%
\delta}{\delta\overline c^{a}} +\partial_{\mu}\frac{\delta}{%
\delta\Omega^{a}_{\mu}}\;,  \nonumber \\
\cr\Bigl\{\mathcal{G}^{a}\;,\;\mathcal{B}_{\Sigma}\Bigr\}&=&\mathcal{W}%
^{a}\;,  \nonumber \\
\cr\Bigl[\mathcal{D}_{\Sigma}\;,\;\mathcal{B}_{\Sigma}\Bigr]&=&0\;,
\nonumber \\
\cr\left[\frac{\delta}{\delta\overline\phi^{a}_{I}} -\partial_{\mu}\frac{%
\delta}{\delta\overline M^{a}_{\mu I}}\;,\, \mathcal{B}_{\Sigma}\right]&=&%
\frac{\delta}{\delta\overline\omega^{a}_{I}} +\partial_{\mu}\frac{\delta}{%
\delta\overline N^{a}_{\mu I}} -gf^{abc}M^{b}_{\mu I}\frac{\delta}{%
\delta\Omega^{c}_{\mu}}\;,  \nonumber \\
\cr\left\{\frac{\delta}{\delta\omega^{a}_{I}} +\partial_{\mu}\frac{\delta}{%
\delta N^{a}_{\mu I}} +gf^{abc}\overline\omega^{b}_{I}\frac{\delta}{\delta
b^{c}}\;,\;\mathcal{B}_{\Sigma}\right\}&=&\frac{\delta\Sigma}{%
\delta\phi^{a}_{I}} -\partial_{\mu}\frac{\delta\Sigma}{\delta M^{a}_{\mu I}}
\nonumber \\
&+&gf^{abc}\biggl(\overline\phi^{b}_{I}\frac{\delta\Sigma}{\delta b^{c}}
+\overline\omega^{b}_{I}\frac{\delta\Sigma}{\delta\overline c^{c}}
+\overline N^{b}_{\mu I}\frac{\delta\Sigma}{\delta\Omega^{c}_{\mu}}\biggr)\;,
\nonumber \\
\cr\Bigl\{\mathcal{R}_{IJ}\;,\;\mathcal{B}_{\Sigma}\Bigr\}&=&\mathcal{Q}%
_{IJ}\;,  \nonumber \\
\cr\Bigl[\mathcal{W}_{I}\;,\;\mathcal{B}_{\Sigma}\Bigr]&=&-\mathcal{Q}%
^{\Sigma}_{I}\;,  \nonumber \\
\cr\Bigl\{\mathcal{R}^{(1)}_{ij}\;,\;\mathcal{B}_{\Sigma}\Bigr\}&=&\mathcal{Q%
}_{ij}\;,  \nonumber \\
\cr\Bigl\{\mathcal{R}^{(2)}_{ij}\;,\;\mathcal{B}_{\Sigma}\Bigr\}&=&0\;,
\nonumber \\
\cr\Bigl\{\mathcal{R}^{(3)}_{ij}\;,\;\mathcal{B}_{\Sigma}\Bigr\}= \int
d^{4}\!x\,&&\!\!\!\!\!\!\!\!\!\!\!\!\!\!\!\left(\delta_{ik}\delta_{jl}-%
\delta_{il}\delta_{jk}\right) \left(\overline B^{a}_{k}\frac{\delta}{\delta
B^{a}_{l}} -\overline V_{k\mu\nu}\frac{\delta}{\delta V_{l\mu\nu}}
+\overline Y^{a}_{k}\frac{\delta}{\delta Y^{a}_{l}} \right)\;,  \nonumber \\
\cr\Bigl\{\mathcal{R}^{(4)}_{ij}\;,\;\mathcal{B}_{\Sigma}\Bigr\}= \int
d^{4}\!x\,&&\!\!\!\!\!\!\!\!\!\!\!\!\!\!\!\left(\delta_{ik}\delta_{jl}+%
\delta_{il}\delta_{jk}\right) \left(G^{a}_{k}\frac{\delta}{\delta\overline
G^{a}_{l}} -U_{k\mu\nu}\frac{\delta}{\delta\overline U_{l\mu\nu}} -X^{a}_{k}%
\frac{\delta}{\delta\overline X^{a}_{l}}\right)\;,  \nonumber \\
\cr\Bigl\{\mathcal{R}^{(1)}_{ik}\;,\;\mathcal{R}^{(3)}_{kj}\Bigr\}= -\int
d^{4}\!x\,&&\!\!\!\!\!\!\!\!\!\!\!\!\!\!\!\left(\delta_{ik}\delta_{jl}+%
\delta_{il}\delta_{jk}\right) \left(\overline G^{a}_{k}\frac{\delta}{\delta
G^{a}_{l}} -\overline U_{k\mu\nu}\frac{\delta}{\delta U_{l\mu\nu}}
-\overline X^{a}_{k}\frac{\delta}{\delta X^{a}_{l}}\right)\;,  \nonumber \\
\cr\Bigl\{\mathcal{R}^{(1)}_{ik}\;,\;\mathcal{R}^{(4)}_{kj}\Bigr\}= -\int
d^{4}\!x\,&&\!\!\!\!\!\!\!\!\!\!\!\!\!\!\!\left(\delta_{ik}\delta_{jl}-%
\delta_{il}\delta_{jk}\right) \left(B^{a}_{k}\frac{\delta}{\delta\overline
B^{a}_{l}} -V_{k\mu\nu}\frac{\delta}{\delta\overline V_{l\mu\nu}} +Y^{a}_{k}%
\frac{\delta}{\delta\overline Y^{a}_{l}} \right)\;.  \nonumber \\
\end{eqnarray}

\sect{Characterization of the most general counterterm}

In order to characterize the most general invariant counterterm which can be
freely added to all orders of perturbation theory, we perturb the classical
action $\Sigma $ by adding an arbitrary integrated local polynomial $\Sigma
_{\mathrm{CT}}$ in the fields and external sources of dimension bounded by
four and with zero ghost number, and we require that the perturbed action $%
\left( \Sigma +\eta \Sigma _{\mathrm{CT}}\right) $ satisfies the same Ward
identities as $\Sigma $ to the first order in the perturbation parameter $%
\eta $. Making use of the BRST cohomological results \cite{Piguet:1995er},
we may write that
\begin{equation}
\Sigma _{\mathrm{CT}}=a_{0}\,S_{\mathrm{YM}}+\mathcal{B}_{\Sigma }\Delta
^{(-1)}\;,
\end{equation}
where $\mathcal{B}_{\Sigma }$ is the nilpotent linearized
Slavnov-Taylor operator of eq.\eqref{linearized},
\begin{equation}
\mathcal{B}_{\Sigma }\mathcal{B}_{\Sigma }=0\;.  \label{npb}
\end{equation}
The expression $\Delta ^{(-1)}$ is an integrated polynomial of ghost
number $-1$, in the present case given by
\begin{eqnarray}
\Delta ^{(-1)} &\!\!\!=\!\!\!&\int d^{4}\!x\,\biggl\{ a_{1}(\Omega _{\mu
}^{a}+\partial _{\mu }\overline{c}^{a})A_{\mu
}^{a}+a_{2}\,L^{a}c^{a}+a_{3}\left( \overline{Y}_{i}^{a}B_{i}^{a}-\overline{X%
}_{i}^{a}G_{i}^{a}+Y_{i}^{a}\overline{B}_{i}^{a}-X_{i}^{a}G_{i}^{a}\right)
+a_{4}\,\overline{N}_{\mu I}^{a}\,\partial _{\mu }\phi _{I}^{a}  \nonumber \\
&&+a_{5}\,M_{\mu I}^{a}\,\partial _{\mu }\overline{\omega }_{I}^{a}+a_{6}\,%
\overline{U}_{i\mu \nu }(\partial _{\mu }A_{\nu }^{a})B_{i}^{a}+a_{7}\,%
\overline{U}_{i\mu \nu }A_{\nu }^{a}\,\partial _{\mu
}B_{i}^{a}+a_{8}\,V_{i\mu \nu }(\partial _{\mu }A_{\nu }^{a})\overline{G}%
_{i}^{a}+a_{9}\,V_{i\mu \nu }A_{\nu }^{a}\,\partial _{\mu }\overline{G}%
_{i}^{a}  \nonumber \\
&&+a_{10}\,\overline{\omega }_{I}^{a}\partial ^{2}\phi _{I}^{a}+a_{11}\,%
\overline{G}_{i}^{a}\partial ^{2}B_{i}^{a}+a_{12}\,\zeta \,\overline{U}%
_{i\mu \nu }V_{i\mu \nu }\left( \overline{V}_{j\alpha \beta }V_{j\alpha
\beta }-\overline{U}_{j\alpha \beta }U_{j\alpha \beta }\right) +a_{13}\,\chi
_{1}\,\overline{U}_{i\mu \nu }\partial ^{2}V_{i\mu \nu }  \nonumber \\
&&+a_{14}\,\chi _{2}\,\overline{U}_{i\mu \nu }\partial _{\mu }\partial
_{\alpha }V_{i\nu \alpha }+a_{15}\,\overline{\omega }_{I}^{a}\phi
_{I}^{a}\left( \overline{V}_{i\mu \nu }V_{i\mu \nu }-\overline{U}_{i\mu \nu
}U_{i\mu \nu }\right) +a_{16}\,\lambda _{1}\,\overline{G}_{i}^{a}B_{i}^{a}%
\left( \overline{V}_{j\mu \nu }V_{j\mu \nu }-\overline{U}_{j\mu \nu }U_{j\mu
\nu }\right)   \nonumber \\
&&+a_{17}\,\lambda _{3}\,\left( \overline{G}_{i}^{a}G_{j}^{a}V_{i\mu \nu }%
\overline{U}_{j\mu \nu }+\overline{G}_{i}^{a}B_{j}^{a}V_{i\mu \nu }\overline{%
V}_{j\mu \nu }-\frac{1}{2}B_{i}^{a}B_{j}^{a}\overline{U}_{i\mu \nu }%
\overline{V}_{j\mu \nu }+\frac{1}{2}B_{i}^{a}G_{j}^{a}\overline{U}_{i\mu \nu
}\overline{U}_{j\mu \nu }\right.   \nonumber \\
&&\left. -\frac{1}{2}\overline{G}_{i}^{a}\overline{B}_{j}^{a}V_{i\mu \nu
}V_{j\mu \nu }+\frac{1}{2}\overline{G}_{i}^{a}\overline{G}_{j}^{a}V_{i\mu
\nu }U_{j\mu \nu }\right) +a_{18}\,\frac{\lambda ^{abcd}+\mathcal{N}^{abcd}}{%
16}\,\overline{G}_{i}^{a}B_{i}^{b}\left( \overline{B}_{j}^{c}B_{j}^{d}-%
\overline{G}_{j}^{c}G_{j}^{d}\right)   \nonumber \\
&&+a_{19}\,\chi \,\overline{N}_{\mu I}^{a}M_{\mu I}^{a}+a_{20}\,gf^{abc}%
\overline{N}_{\mu I}^{a}\,\phi _{I}^{b}A_{\mu }^{c}+a_{21}\,gf^{abc}M_{\mu
I}^{a}\,\overline{\omega }_{I}^{b}A_{\mu }^{c}+a_{22}\,gf^{abc}\overline{%
\omega }_{I}^{a}A_{\mu }^{c}\,\partial _{\mu }\phi _{I}^{b}  \nonumber \\
&&+a_{23}\,gf^{abc}\overline{\omega }_{I}^{a}(\partial _{\mu }A_{\mu
}^{c})\phi _{I}^{b}+a_{24}\,gf^{abc}\overline{G}_{i}^{a}A_{\mu
}^{c}\,\partial _{\mu }B_{i}^{b}+a_{25}\,gf^{abc}\overline{G}%
_{i}^{a}(\partial _{\mu }A_{\mu }^{c})B_{i}^{b}  \nonumber \\
&&+a_{26}\,gf^{abc}\overline{U}_{i\mu \nu }A_{\mu }^{b}A_{\nu
}^{c}B_{i}^{a}+a_{27}\,gf^{abc}V_{i\mu \nu }A_{\mu }^{b}A_{\nu }^{c}%
\overline{G}_{i}^{a}+\tilde{\lambda}^{abcd}\overline{G}_{i}^{a}B_{i}^{b}A_{%
\mu }^{c}A_{\mu }^{d}\biggr\}\;.
\end{eqnarray}
Due to the Ward identities given in section 4, the counterterm
$\Sigma _{\mathrm{CT}}$ must obey the following constraints
\begin{eqnarray}
&&{\cal B}_{\Sigma}\Sigma_{\mathrm{CT}}=0\;,\quad\,\, {\cal
D}_{\Sigma}\Sigma_{\mathrm{CT}}=0\;,\quad\;\;\;\;\;\;\;\; {\cal
G}^{a}\Sigma_{\mathrm{CT}}=0\;,\qquad {\cal
W}^{a}\Sigma_{\mathrm{CT}}=0\;,\nonumber\cr &&{\cal
Q}_{IJ}\Sigma_{\mathrm{CT}}=0\;,\quad {\cal
R}_{IJ}\Sigma_{\mathrm{CT}}=0\;,\quad\;\;\;\;\;\, {\cal
W}_{I}\Sigma_{\mathrm{CT}}=0\;,\quad {\cal
Q}^{\Sigma}_{I}\Sigma_{\mathrm{CT}}=0\;,\nonumber\cr &&{\cal
Q}_{ij}\Sigma_{\mathrm{CT}}=0\;,\quad\, {\cal
R}^{(1,2,3,4)}_{ij}\Sigma_{\mathrm{CT}}=0\;,\quad \{{\cal
R}^{(3,4)}_{ij},{\cal B}_{\Sigma}\}\Sigma_{\mathrm{CT}}=0\;,\quad
\{{\cal R}^{(1)}_{ik},{\cal
R}^{(3,4)}_{kj}\}\Sigma_{\mathrm{CT}}=0\;,\nonumber\cr\cr
&&\frac{\delta\Sigma_{\mathrm{CT}}}{\delta b^{a}}=0\;,\quad
\frac{\delta\Sigma_{\mathrm{CT}}}{\delta \overline c^{a}}
+\partial_{\mu}\frac{\delta\Sigma_{\mathrm{CT}}}{\delta\Omega^{a}_{\mu}}=0\;,\quad
\frac{\delta\Sigma_{\mathrm{CT}}}{\delta \overline\phi^{a}_{I}}
-\partial_{\mu}\frac{\delta\Sigma_{\mathrm{CT}}}{\delta\overline
M^{a}_{\mu I}}=0\;,\quad
\frac{\delta\Sigma_{\mathrm{CT}}}{\delta\omega^{a}_{I}}
+\partial_{\mu}\frac{\delta\Sigma_{\mathrm{CT}}}{\delta N^{a}_{\mu
I}}=0\;,\nonumber\cr
&&\frac{\delta\Sigma_{\mathrm{CT}}}{\delta\phi^{a}_{I}}
-\partial_{\mu}\frac{\delta\Sigma_{\mathrm{CT}}}{\delta M^{a}_{\mu
I}} +gf^{abc}\biggl(
\overline\omega^{b}_{I}\frac{\delta\Sigma_{\mathrm{CT}}}{\delta\overline
c^{c}}
+\overline N^{b}_{\mu I}\frac{\delta\Sigma_{\mathrm{CT}}}{\delta\Omega^{c}_{\mu}} \biggr)=0\;,\nonumber\\
&&\frac{\delta\Sigma_{\mathrm{CT}}}{\delta\overline\omega^{a}_{I}}
+\partial_{\mu}\frac{\delta\Sigma_{\mathrm{CT}}}{\delta\overline
N^{a}_{\mu I}} -gf^{abc}M^{b}_{\mu
I}\frac{\delta\Sigma_{\mathrm{CT}}}{\delta\Omega^{c}_{\mu}}=0\;.
\label{constraints}
\end{eqnarray}
By applying the constraints \eqref{constraints}, one can show that
\begin{equation}
\tilde{\lambda}^{abcd}=a_{11}\,g^{2}f^{ace}f^{edb}\;,
\end{equation}
and
\begin{eqnarray}
\Sigma _{\mathrm{CT}} &\!\!\!=\!\!\!&a_{0}\,S_{\mathrm{YM}}+\int d^{4}\!x\,%
\biggl\{a_{1}\biggl[A_{\mu }^{a}\frac{\delta S_{\mathrm{YM}}}{\delta A_{\mu
}^{a}}+(\Omega _{\mu }^{a}+\partial _{\mu }\overline{c}^{a})\partial _{\mu
}c^{a}+\overline{\phi }_{I}^{a}\partial ^{2}\phi _{I}^{a}-\overline{\omega }%
_{I}^{a}\partial ^{2}\omega _{I}^{a}-gf^{abc}\overline{\omega }%
_{I}^{a}\partial _{\mu }(\phi _{I}^{b}\partial _{\mu }c^{c})  \nonumber \\
&&+\overline{M}_{\mu I}^{a}\partial _{\mu }\phi _{I}^{a}+\overline{N}_{\mu
I}^{a}(\partial _{\mu }\omega _{I}^{a}+gf^{abc}\phi _{I}^{b}\partial _{\mu
}c^{c})-N_{\mu I}^{a}\partial _{\mu }\overline{\omega }_{I}^{a}+M_{\mu
I}^{a}(\partial _{\mu }\overline{\phi }_{I}^{a}+gf^{abc}\overline{\omega }%
_{I}^{b}\partial _{\mu }c^{c})-(\overline{M}_{\mu }^{a}M_{\mu }^{a}
\nonumber \\
&&-\overline{N}_{\mu }^{a}N_{\mu }^{a})\biggr] +(2a_{2}+a_{3})(\overline{B}%
_{i}^{a}\partial ^{2}B_{i}^{a}-\overline{G}_{i}^{a}\partial
^{2}G_{i}^{a})-(a_{1}+2a_{2}+a_{3})gf^{abc}[\overline{B}_{i}^{a}(\partial
_{\mu }A_{\mu }^{b}+2A_{\mu }^{b}\partial _{\mu })B_{i}^{c}  \nonumber \\
&&-\overline{G}_{i}^{a}(\partial _{\mu }A_{\mu }^{b}+2A_{\mu }^{b}\partial
_{\mu })G_{i}^{c}]+(2a_{1}+2a_{2}+a_{3})g^{2}f^{abd}f^{bce}(\overline{B}%
_{i}^{a}B_{i}^{c}-\overline{G}_{i}^{a}G_{i}^{c})A_{\mu }^{d}A_{\mu }^{e}
\nonumber
\end{eqnarray}
\begin{eqnarray}
&&+[(a_{1}+a_{2}+a_{4})2\partial _{\mu }A_{\nu
}^{a}+(2a_{1}+a_{2}+a_{4})gf^{abc}A_{\mu }^{b}A_{\nu }^{c}](\overline{U}%
_{i\mu \nu }G_{i}^{a}+V_{i\mu \nu }\overline{B}_{i}^{a}+U_{i\mu \nu }%
\overline{G}_{i}^{a}-\overline{V}_{i\mu \nu }B_{i}^{a})  \nonumber \\
&&+(4a_{2}+a_{5})\frac{\lambda ^{abcd}+\mathcal{N}^{abcd}}{16}(\overline{B}%
_{i}^{a}B_{i}^{b}-\overline{G}_{i}^{a}G_{i}^{b})(\overline{B}%
_{j}^{c}B_{j}^{d}-\overline{G}_{j}^{c}G_{j}^{d})+(2a_{2}+a_{6})\lambda _{1}(%
\overline{B}_{i}^{a}B_{i}^{a}-\overline{G}_{i}^{a}G_{i}^{a})  \nonumber \\
&&\times (\overline{V}_{i\mu \nu }V_{i\mu \nu }-\overline{U}_{i\mu \nu
}U_{i\mu \nu })+(2a_{2}+a_{7})\lambda _{3}\Bigl(\overline{B}%
_{i}^{a}G_{j}^{a}V_{i\mu \nu }\overline{U}_{j\mu \nu }+\overline{G}%
_{i}^{a}G_{j}^{a}U_{i\mu \nu }\overline{U}_{j\mu \nu }+\overline{B}%
_{i}^{a}B_{j}^{a}V_{i\mu \nu }\overline{V}_{j\mu \nu }  \nonumber \\
&&-\overline{G}_{i}^{a}B_{j}^{a}V_{j\mu \nu }U_{i\mu \nu }-G_{i}^{a}B_{j}^{a}%
\overline{U}_{i\mu \nu }\overline{V}_{j\mu \nu }+\overline{G}_{i}^{a}%
\overline{B}_{j}^{a}U_{i\mu \nu }V_{j\mu \nu }-\frac{1}{2}B_{i}^{a}B_{j}^{a}%
\overline{V}_{i\mu \nu }\overline{V}_{j\mu \nu }+\frac{1}{2}%
G_{i}^{a}G_{j}^{a}\overline{U}_{i\mu \nu }\overline{U}_{j\mu \nu }  \nonumber
\\
&&-\frac{1}{2}\overline{B}_{i}^{a}\overline{B}_{j}^{a}V_{i\mu \nu }V_{j\mu
\nu }+\frac{1}{2}\overline{G}_{i}^{a}\overline{G}_{j}^{a}U_{i\mu \nu
}U_{j\mu \nu }\Bigr)+a_{8}\,\zeta (\overline{U}_{i\mu \nu }U_{i\mu \nu }%
\overline{U}_{j\alpha \beta }U_{j\alpha \beta }+\overline{V}_{i\mu \nu
}V_{i\mu \nu }\overline{V}_{j\alpha \beta }V_{j\alpha \beta }  \nonumber \\
&&-2\overline{U}_{i\mu \nu }U_{i\mu \nu }\overline{V}_{j\alpha \beta
}V_{j\alpha \beta })+a_{9}\,\chi _{1}(\overline{V}_{i\mu \nu }\partial
^{2}V_{i\mu \nu }-\overline{U}_{i\mu \nu }\partial ^{2}U_{i\mu \nu })
\nonumber \\
&&+a_{10}\,\chi _{2}(\overline{V}_{i\mu \nu }\partial _{\mu }\partial
_{\alpha }V_{i\nu \alpha }-\overline{U}_{i\mu \nu }\partial _{\mu }\partial
_{\alpha }U_{i\nu \alpha })\biggl\}\;,
\end{eqnarray}
where we renamed the coefficients $a_{n}$ as
\begin{equation}
\begin{tabular}{rcl|rcl|rcl}
$a_{3}$ & $\!\!\to \!\!$ & $a_{2},$ & $a_{18}$ & $\!\!\to \!\!$ & $a_{5},$ &
$a_{12}$ & $\!\!\to \!\!$ & $a_{8},$ \\
$a_{11}$ & $\!\!\to \!\!$ & $a_{3},$ & $a_{16}$ & $\!\!\to \!\!$ & $a_{6},$
& $a_{13}$ & $\!\!\to \!\!$ & $a_{9},$ \\
$a_{6}$ & $\!\!\to \!\!$ & $-2a_{4},$ & $a_{17}$ & $\!\!\to \!\!$ & $a_{7},$
& $a_{14}$ & $\!\!\to \!\!$ & $a_{10}.$%
\end{tabular}
\end{equation}

\sect{Stability of the action at the quantum level and
renormalization factors}

As a final step, we must show that the most general counterterm $\Sigma _{%
\mathrm{CT}}$ can be reabsorbed by means of a multiplicative renormalization
of the parameters, fields, and sources already present in the starting
action $\Sigma $. Taking
\begin{eqnarray}
\psi _{0} &=&Z_{\psi }^{1/2}\,\psi \;,  \nonumber \\
J_{0} &=&Z_{J}\,J\;,  \nonumber \\
\xi _{0} &=&Z_{\xi }\,\xi \;,  \nonumber \\
\lambda _{0}^{abcd} &=&Z_{\lambda }\,\lambda ^{abcd}+\mathcal{Z}^{abcd}\;,
\end{eqnarray}
where
\begin{eqnarray}
\psi  &=&\{A,b,c,\overline{c},\phi ,\overline{\phi },\omega ,\overline{%
\omega },B,\overline{B},G,\overline{G}\}\;,  \nonumber \\
J &=&\{\Omega ,L,M,\overline{M},N,\overline{N},U,\overline{U},V,\overline{V}%
,X,\overline{X},Y,\overline{Y}\}\;,  \nonumber \\
\xi  &=&\{g,\chi ,\chi _{1},\chi _{2},\zeta ,\lambda _{1},\lambda _{3}\}\;,
\end{eqnarray}
we must show that
\begin{equation}
\Sigma (\psi _{0},J_{0},\xi _{0})=\Sigma (\psi ,J,\xi )+\eta \,\Sigma _{%
\mathrm{CT}}(\psi ,J,\xi )+O(\eta ^{2})\;.
\end{equation}
After some algebra, for the renormalization factors $\left\{
Z\right\} $ we obtain
\begin{eqnarray}
Z_{b} &\!\!=\!\!&Z_{A}^{-1}\;,\quad Z_{c}=Z_{\overline{c}%
}=Z_{g}^{-1}Z_{A}^{-1/2}\;,  \nonumber \\
Z_{\phi } &\!\!=\!\!&Z_{\overline{\phi }}=Z_{g}^{-1}Z_{A}^{-1/2}\;,\quad Z_{%
\overline{\omega }}=Z_{g}^{-2}\;,\quad Z_{\omega }=Z_{A}^{-1}\;,  \nonumber
\\
Z_{M} &\!\!=\!\!&Z_{\overline{M}}=Z_{\phi }^{1/2}=Z_{g}^{-1/2}Z_{A}^{-1/4}\;,
\nonumber \\
Z_{\overline{N}} &\!\!=\!\!&Z_{\overline{\omega }}^{1/2}=Z_{g}^{-1}\;,\quad
Z_{N}=Z_{\omega }^{1/2}=Z_{A}^{-1/2}\;,  \nonumber \\
Z_{\Omega }
&\!\!=\!\!&Z_{c}^{1/2}=Z_{g}^{-1/2}Z_{A}^{-1/4}\;,Z_{L}=Z_{A}^{1/2}\;,
\nonumber \\
Z_{X} &\!\!=\!\!&Z_{\overline{X}}=Z_{Y}=Z_{\overline{Y}%
}=Z_{g}^{-1/2}Z_{A}^{1/4}Z_{B}^{-1/2}\;,  \nonumber \\
Z_{A} &\!\!=\!\!&1+\eta (a_{0}+2a_{1})\;,\quad Z_{g}=1-\eta \frac{a_{0}}{2}%
\;,  \nonumber \\
Z_{B} &\!\!=\!\!&Z_{\overline{B}}=Z_{G}=Z_{\overline{G}}=1+\eta
(2a_{2}+a_{3})\;,  \nonumber \\
Z_{V} &\!\!=\!\!&Z_{\overline{V}}=Z_{U}=Z_{\overline{U}}=1-\eta \Bigl(\frac{%
a_{0}}{2}+\frac{a_{3}}{2}-a_{4}\Bigr)\;,  \nonumber \\
Z_{\lambda } &\!\!=\!\!&1+\eta (a_{5}-2a_{3})\;,\quad \mathcal{Z}%
^{abcd}=\,\eta (4a_{2}+a_{5})\mathcal{N}^{abcd}\;,  \nonumber \\
Z_{\lambda _{1}} &\!\!=\!\!&1+\eta (a_{0}-2a_{4}+a_{6})\;,\quad Z_{\lambda
_{3}}=1+\eta (a_{0}-2a_{4}+a_{7})\;,  \nonumber \\
Z_{\zeta } &\!\!=\!\!&1+\eta (2a_{0}+2a_{3}-4a_{4}-a_{8})\;,\quad Z_{\chi
_{1}}=1+\eta (a_{0}+a_{3}-2a_{4}+a_{9})\;,  \nonumber \\
Z_{\chi _{2}} &\!\!=\!\!&1+\eta (a_{0}+a_{3}-2a_{4}+a_{10})\;,
\end{eqnarray}
hereby confirming the renormalizability to all orders of perturbation theory
of the action \eqref{sigma}. We draw attention to the fact that the
renormalization of the quartic tensor coupling $\lambda ^{abcd}$ involves an
additional additive part given by $\mathcal{Z}^{abcd}$ \cite{Capri:2006ne}.
As a consequence, $\lambda ^{abcd}=0$ is a not a fixed point of the model.
This originates from the fact that the interactions proportional to the
other coupling $g^{2}$ reintroduce by quantum effects the tensor coupling $%
\propto \left( \overline{B}_{\mu \nu }^{a}B_{\mu \nu }^{b}-\overline{G}_{\mu
\nu }^{a}G_{\mu \nu }^{b}\right) \left( \overline{B}_{\rho \sigma
}^{c}B_{\rho \sigma }^{d}-\overline{G}_{\rho \sigma }^{c}G_{\rho \sigma
}^{d}\right) $. This is nicely reflected in the renormalization group
function of $\lambda ^{abcd}$, that was calculated at one loop order in \cite
{Capri:2006ne} using dimensional regularization in $d=4-2\varepsilon $
dimensions and the $\overline{\mbox{MS}}$ scheme,
\begin{eqnarray}
\mu \frac{\partial }{\partial \mu }\lambda ^{abcd} &=&-2\varepsilon \lambda
^{abcd}+\left[ \frac{1}{4}\left( \lambda ^{abpq}\lambda ^{cpdq}+\lambda
^{apbq}\lambda ^{cdpq}+\lambda ^{apcq}\lambda ^{bpdq}+\lambda ^{apdq}\lambda
^{bpcq}\right) \right.   \nonumber  \label{betaabcdbis} \\
&&\left. -~12C_{A}\lambda
^{abcd}a~+~8C_{A}f^{abp}f^{cdp}a^{2}~+~16C_{A}f^{adp}f^{bcp}a^{2}~+~96d_{A}^{abcd}a^{2}\right] \;,
\end{eqnarray}
where $a=\frac{g^{2}}{16\pi ^{2}}$ and $d_{A}^{abcd}$ is the totally
symmetric rank four tensor defined by $d_{A}^{abcd}=\mbox{Tr}\left(
T_{A}^{a}T_{A}^{(b}T_{A}^{c}T_{A}^{d)}\right) $. Clearly, we have
\begin{eqnarray} \mu \frac{\partial }{\partial \mu }\lambda ^{abcd}\neq 0\mbox{
for
    }\lambda ^{abcd}=0\;.
\end{eqnarray}

\subsection{{The physical action and some of its properties} }

We have shown the renormalizability of the complete action \eqref{sigma}. In
particular, since the renormalizability holds for all possible values of the
sources, we have also proven it in the case that the external source part, %
\eqref{ext}, is zero, while the other sources attain their physical values
dictated by \eqref{phys1} and \eqref{phys2}, yielding the complete physical
action
\begin{eqnarray}
S_{\mathrm{physical}} &=&\int d^{4}x\left( \frac{1}{4}F_{\mu \nu }^{a}F_{\mu
\nu }^{a}\right) +\int d^{4}x\;\left( b^{a}\partial _{\mu }A_{\mu }^{a}+%
\overline{c}^{a}\partial _{\mu }D_{\mu }^{ab}c^{b}\right)   \nonumber
\label{final} \\
&+&\int d^{4}x\left( -\overline{\varphi }_{\mu }^{ac}\partial _{\nu }\left(
\partial _{\nu }\varphi _{\mu }^{ac}+gf^{abm}A_{\nu }^{b}\varphi _{\mu
}^{mc}\right) +\overline{\omega }_{\mu }^{ac}\partial _{\nu }\left( \partial
_{\nu }\omega _{\mu }^{ac}+gf^{abm}A_{\nu }^{b}\omega _{\mu }^{mc}\right)
\right.   \nonumber \\
&&+\left. \overline{\omega }_{\mu }^{ac}\partial _{\nu }\Bigl(gf^{abd}\phi
_{\mu }^{bc}\,D_{\nu }^{de}c^{e}\Bigr)\right)   \nonumber \\
&+&\int d^{4}x\left[ \gamma ^{2}gf^{abc}A_{\mu }^{b}\varphi _{\mu
}^{ac}-\gamma ^{2}gf^{abc}A_{\mu }^{b}\overline{\varphi }_{\mu
}^{ac}-4\left( N^{2}-1\right) \gamma ^{4}\right]   \nonumber \\
&+&\int d^{4}x\left( \frac{im}{4}(B-\overline{B})_{\mu \nu }^{a}F_{\mu \nu
}^{a}+\frac{1}{4}\left( \overline{B}_{\mu \nu }^{a}D_{\sigma }^{ab}D_{\sigma
}^{bc}B_{\mu \nu }^{c}-\overline{G}_{\mu \nu }^{a}D_{\sigma }^{ab}D_{\sigma
}^{bc}G_{\mu \nu }^{c}\right) \right)   \nonumber \\
&+&\int d^{4}x\left( -\frac{3}{8}m^{2}\lambda _{1}\left( \overline{B}_{\mu
\nu }^{a}B_{\mu \nu }^{a}-\overline{G}_{\mu \nu }^{a}G_{\mu \nu }^{a}\right)
+m^{2}\frac{\lambda _{3}}{32}\left( \overline{B}_{\mu \nu }^{a}-B_{\mu \nu
}^{a}\right) ^{2}\right)   \nonumber \\
&+&\int d^{4}x\left( \frac{\lambda ^{abcd}}{16}\left( \overline{B}_{\mu \nu
}^{a}B_{\mu \nu }^{b}-\overline{G}_{\mu \nu }^{a}G_{\mu \nu }^{b}\right)
\left( \overline{B}_{\rho \sigma }^{c}B_{\rho \sigma }^{d}-\overline{G}%
_{\rho \sigma }^{c}G_{\rho \sigma }^{d}\right) \right) -\int
d^{4}x\left( \frac{9}{4}\zeta m^{4}\right)\;. \nonumber\\
\end{eqnarray}
Apparently, the symmetry content of the action \eqref{sigma} given in
section 3 is sufficiently powerful to avoid mixing between the Zwanziger
fields/sources on one hand and the mass related fields/sources on the other
hand.

\noindent The term $\propto \zeta m^4$ in the final action \eqref{final} is
irrelevant for the renormalization of Green functions, but it becomes
important when one looks at the renormalization of the vacuum energy $E(m)$.
The parameter $\zeta$ is the so-called LCO parameter, and its value ought to
be fixed by requiring a homogenous linear renormalization group equation for
$E(m)$, whereby $\zeta$ is made a function of the available couplings. This
point is however beyond the scope of this paper, the interested reader is
kindly referred to e.g. \cite
{Verschelde:1995jj,Verschelde:2001ia,Dudal:2004rx,Knecht:2001cc} for more
details. It is a remarkable feature of the Zwanziger action that there is no
need for such a LCO parameter in front of the $\gamma^4$-term in the action %
\eqref{final} \cite{Zwanziger:1992qr,Maggiore:1993wq,Dudal:2005na}.

\noindent If we make abstract of the Gribov-Zwanziger part for the moment,
we established a ``supersymmetry'' for the action $S_{\mathrm{physical}%
}^{m=0,\gamma=0}$, generated by \cite{Capri:2006ne}
\begin{eqnarray}  \label{ss}
\delta_s B_{\mu\nu}^a &=& G_{\mu\nu}^a \;,\qquad \delta_s G_{\mu\nu}^a =0 \;,
\nonumber \\
\delta_s \overline{G}_{\mu\nu}^a &=& \overline{B}_{\mu\nu}^a \;, \qquad
\delta_s \overline{B}_{\mu\nu}^a = 0 \;,  \nonumber \\
\delta_s \Psi&=&0 \text{ for all other fields }\Psi\;,  \nonumber \\
\delta_s^2&=&0\;,  \nonumber \\
\delta_s \left(S_\mathrm{physical}^{m=0,\gamma=0}\right)&=&0\;.
\end{eqnarray}
We used this symmetry in \cite{Capri:2006ne} to show that the massless
version of our gauge model is equivalent with Yang-Mills ordinary gauge
theories, despite the extra (quartic) interactions between the fields $%
B_{\mu\nu}^a$, $\overline{B}_{\mu\nu}^a$, $G_{\mu\nu}^a$ and $\overline{G}%
_{\mu\nu}^a$. A completely similar $\delta_s$-cohomological argument as
presented in \cite{Capri:2006ne} can be used here to actually prove that the
action $S_{\mathrm{physical}}^{m=0}$ and the original Gribov-Zwanziger
action give rise to the \emph{same} Green functions at any order of
perturbation theory when we restricts ourselves to those functionals built
from fields in the original Gribov-Zwanziger action, meaning that the
quartic coupling $\lambda^{abcd}$ cancels out from the final results.

\noindent The combination of the previous result and the already mentioned
absence of mixing, also implies that the already known renormalization group
functions and relations for the Gribov-Zwanziger action \cite
{Zwanziger:1992qr,Dudal:2005na,Maggiore:1993wq} and massive gauge model \cite
{Capri:2005dy,Capri:2006ne} remain valid when both are combined into one
action, at least whenever massless renormalization schemes like the $%
\overline{\mbox{MS}}$ one are employed.

\noindent When the sources are set equal to their physical values %
\eqref{phys1} and \eqref{phys2} in order to obtain the action $S_{\mathrm{%
physical}}$, the BRST symmetry \eqref{brst} is however broken. It is worth
having a somewhat more detailed look at this.

\sect{The breaking of the Slavnov-Taylor identity scrutinized }

\subsection{The case of the massive gauge model without the Gribov
restriction}

In order to avoid too lengthy expressions, we shall momentarily skip
the Gribov restriction, and concentrate on the massive gauge model
already studied in earlier papers \cite {Capri:2005dy,Capri:2006ne}.

\noindent Let $\widetilde{\Sigma }$ thus be the complete action given by
\begin{eqnarray}  \label{a1}
\widetilde{\Sigma} &=&S_{\mathrm{YM}} +\int d^{4}\!x\,\biggl\{%
b^{a}\,\partial_{\mu}A^{a}_{\mu} +\overline
c^{a}\partial_{\mu}D^{ab}_{\mu}c^{b} +\overline
B^{a}_{i}D^{ab}_{\mu}D^{bc}_{\mu}B^{c}_{i} -\overline
G^{a}_{i}D^{ab}_{\mu}D^{bc}_{\mu}G^{c}_{i}  \nonumber \\
&& +F^{a}_{\mu\nu}\bigl(\overline U_{i\mu\nu}G^{a}_{i} +V_{i\mu\nu}\overline
B^{a}_{i} -\overline V_{i\mu\nu}B^{a}_{i} +U_{i\mu\nu}\overline G^{a}_{i}%
\bigr) +\lambda_{1}(\overline B^{a}_{i}B^{a}_{i}-\overline
G^{a}_{i}G^{a}_{i}) (\overline V_{j\mu\nu}V_{j\mu\nu}-\overline
U_{j\mu\nu}U_{j\mu\nu})  \nonumber \\
&&+\frac{\lambda^{abcd}}{16}(\overline B^{a}_{i}B^{b}_{i}-\overline
G^{a}_{i}G^{b}_{i}) (\overline B^{c}_{j}B^{d}_{j}-\overline
G^{c}_{j}G^{d}_{j}) +\lambda_{3}\Bigl(\overline
B^{a}_{i}G^{a}_{j}V_{i\mu\nu}\overline U_{j\mu\nu} +\overline
G^{a}_{i}G^{a}_{j}U_{i\mu\nu}\overline U_{j\mu\nu}  \nonumber \\
&& +\overline B^{a}_{i}B^{a}_{j}V_{i\mu\nu}\overline V_{j\mu\nu} -\overline
G^{a}_{i}B^{a}_{j}V_{j\mu\nu}U_{i\mu\nu} -G^{a}_{i}B^{a}_{j}\overline
U_{i\mu\nu}\overline V_{j\mu\nu} +\overline G^{a}_{i}\overline
B^{a}_{j}U_{i\mu\nu}V_{j\mu\nu} -\frac{1}{2}B^{a}_{i}B^{a}_{j}\overline
V_{i\mu\nu}\overline V_{j\mu\nu}  \nonumber \\
&&+\frac{1}{2}G^{a}_{i}G^{a}_{j}\overline U_{i\mu\nu}\overline U_{j\mu\nu} -%
\frac{1}{2}\overline B^{a}_{i}\overline B^{a}_{j}V_{i\mu\nu}V_{j\mu\nu} +%
\frac{1}{2}\overline G^{a}_{i}\overline G^{a}_{j}U_{i\mu\nu}U_{j\mu\nu}\Bigr)
\nonumber \\
&& +\chi_{1}(\overline V_{i\mu\nu}\partial^{2}V_{i\mu\nu} -\overline
U_{i\mu\nu}\partial^{2}U_{i\mu\nu}) +\chi_{2}(\overline
V_{i\mu\nu}\partial_{\mu}\partial_{\alpha}V_{i\nu\alpha} -\overline
U_{i\mu\nu}\partial_{\mu}\partial_{\alpha}U_{i\nu\alpha})  \nonumber \\
&&-\zeta(\overline U_{i\mu\nu}U_{i\mu\nu}\overline
U_{j\alpha\beta}U_{j\alpha\beta} +\overline V_{i\mu\nu}V_{i\mu\nu}\overline
V_{j\alpha\beta}V _{j\alpha\beta} -2\overline
U_{i\mu\nu}U_{i\mu\nu}\overline V_{j\alpha\beta}V_{j\alpha\beta})
-\Omega^{a}_{\mu}D^{ab}_{\mu}c^{b}  \nonumber \\
&&+\frac{g}{2}f^{abc}L^{a}c^{b}c^{c}+gf^{abc}\overline
Y^{a}_{i}c^{b}B^{c}_{i} +gf^{abc}Y^{a}_{i}c^{b}\overline B^{c}_{i}
+gf^{abc}\overline X^{a}_{i}c^{b}G^{c}_{i} +gf^{abc}X^{a}_{i}c^{b}\overline
G^{c}_{i}\biggr\}\;.
\end{eqnarray}
This action $\widetilde{\Sigma }$ obeys the Slavnov-Taylor identity
\begin{equation}
\widetilde{\mathcal{S}}(\widetilde{\Sigma })=0\;,  \label{a2}
\end{equation}
with
\begin{eqnarray}
\widetilde{\mathcal{S}}(\widetilde{\Sigma }) &=&\int d^{4}x\left[ \frac{%
\delta \widetilde{\Sigma }}{\delta \Omega _{\mu }^{a}}\frac{\delta
\widetilde{\Sigma }}{\delta A_{\mu }^{a}}+\frac{\delta \widetilde{\Sigma }}{%
\delta L^{a}}\frac{\delta \widetilde{\Sigma }}{\delta c^{a}}+b^{a}\frac{%
\delta \widetilde{\Sigma }}{\delta \overline{c}^{a}}+\left( \frac{\delta
\widetilde{\Sigma }}{\delta \overline{Y}_{i}^{a}}+G_{i}^{a}\right) \frac{%
\delta \widetilde{\Sigma }}{\delta B_{i}^{a}}+\frac{\delta \widetilde{\Sigma
}}{\delta Y_{i}^{a}}\frac{\delta \widetilde{\Sigma }}{\delta \overline{B}%
_{i}^{a}}+\frac{\delta \widetilde{\Sigma }}{\delta \overline{X}_{i}^{a}}%
\frac{\delta \widetilde{\Sigma }}{\delta {G}_{i}^{a}}\right.  \nonumber \\
&+&\left. \left( \frac{\delta \widetilde{\Sigma }}{\delta X_{i}^{a}}+%
\overline{B}_{i}^{a}\right) \frac{\delta \widetilde{\Sigma }}{\delta
\overline{G}_{i}^{a}}+\overline{V}_{i\mu \nu }\frac{\delta \widetilde{\Sigma
}}{\delta \overline{U}_{i\mu \nu }}+U_{i\mu \nu }\frac{\delta \widetilde{%
\Sigma }}{\delta {V}_{i\mu \nu }}-\overline{Y}_{i}^{a}\frac{\delta
\widetilde{\Sigma }}{\delta \overline{X}_{i}^{a}}+X_{i}^{a}\frac{\delta
\widetilde{\Sigma }}{\delta Y_{i}^{a}}\right] \;.  \label{a3}
\end{eqnarray}
\noindent Since the theory is stable and free from anomalies at the quantum
level, we may write down a renormalized 1PI quantum vertex functional \cite
{Piguet:1995er},
\begin{equation}
\widetilde{\Gamma} =\widetilde{\Sigma }+\hbar \widetilde{\Gamma}\;,
^{(1)}+....  \label{a4}
\end{equation}
which fulfills the quantum version of the Slavnov-Taylor identity (\ref{a2}%
), i.e.
\begin{equation}
\widetilde{\mathcal{S}}(\widetilde{\Gamma} )=0\;,  \label{a5}
\end{equation}
\begin{eqnarray}
\widetilde{\mathcal{S}}(\widetilde{\Gamma} ) &=&\int d^{4}x\left[ \frac{%
\delta \widetilde{\Gamma} }{\delta \Omega _{\mu }^{a}}\frac{\delta
\widetilde{\Gamma} }{\delta A_{\mu }^{a}}+\frac{\delta \widetilde{\Gamma} }{%
\delta L^{a}}\frac{\delta \widetilde{\Gamma} }{\delta c^{a}}+b^{a}\frac{%
\delta \widetilde{\Gamma} }{\delta \overline{c}^{a}}+\left( \frac{\delta
\widetilde{\Gamma} }{\delta \overline{Y}_{i}^{a}}+G_{i}^{a}\right) \frac{%
\delta \widetilde{\Gamma} }{\delta B_{i}^{a}}+\frac{\delta \widetilde{\Gamma}
}{\delta Y_{i}^{a}}\frac{\delta \widetilde{\Gamma} }{\delta \overline{B}%
_{i}^{a}}+\frac{\delta \widetilde{\Gamma}}{\delta \overline{X}_{i}^{a}}\frac{%
\delta \widetilde{\Gamma} }{\delta {G}_{i}^{a}}\right.  \nonumber \\
&+&\left. \left( \frac{\delta \widetilde{\Gamma} }{\delta X_{i}^{a}}+%
\overline{B}_{i}^{a}\right) \frac{\delta \widetilde{\Gamma} }{\delta
\overline{G}_{i}^{a}}+\overline{V}_{i\mu \nu }\frac{\delta \widetilde{\Gamma}
}{\delta \overline{U}_{i\mu \nu }}+U_{i\mu \nu }\frac{\delta \widetilde{%
\Gamma} }{\delta {V}_{i\mu \nu }}-\overline{Y}_{i}^{a}\frac{\delta
\widetilde{\Gamma} }{\delta \overline{X}_{i}^{a}}+X_{i}^{a}\frac{\delta
\widetilde{\Gamma} }{\delta Y_{i}^{a}}\right] \;.  \label{a6}
\end{eqnarray}
\noindent Let us now analyse the quantum properties of the model when the
sources attain their physical values \eqref{phys2}. First of all, let us
give a look at the classical action $\widetilde{\Sigma }_{ph},$ obtained by $%
\widetilde{\Sigma }$ by setting the sources to their physical values, namely
\begin{equation}
\widetilde{\Sigma} _{\mathrm{ph}}=\left. \widetilde{\Sigma }\right| _{%
\mathrm{physical\,value\,of\, }\left( V_{i\mu \nu },\overline{V}_{i\mu \nu
},U_{i\mu \nu },\overline{U}_{i\mu \nu }\right) }\;,  \label{n1}
\end{equation}
or explicitly
\begin{eqnarray}  \label{finalzondergribov}
\widetilde{\Sigma }_{\mathrm{ph}}&=&\int d^4x\left(\frac{1}{4}F_{\mu \nu
}^{a}F_{\mu \nu }^{a}\right)+\int d^{4}x\;\left( b^{a}\partial_\mu A_\mu^{a}+%
\overline{c}^{a}\partial _{\mu }D_{\mu }^{ab}c ^{b}\right)  \nonumber \\
&+&\int d^4x\left(\frac{im}{4}(B-\overline{B})_{\mu\nu}^aF_{\mu\nu}^a +\frac{%
1}{4}\left( \overline{B}_{\mu \nu }^{a}D_{\sigma }^{ab}D_{\sigma
}^{bc}B_{\mu \nu }^{c}-\overline{G}_{\mu \nu }^{a}D_{\sigma }^{ab}D_{\sigma
}^{bc}G_{\mu \nu }^{c}\right)\right)  \nonumber \\
&+&\int d^4x\left(-\frac{3}{8}m^{2}\lambda _{1}\left( \overline{B}_{\mu \nu
}^{a}B_{\mu \nu }^{a}-\overline{G}_{\mu \nu }^{a}G_{\mu \nu }^{a}\right)
+m^{2}\frac{\lambda _{3}}{32}\left( \overline{B}_{\mu \nu }^{a}-B_{\mu \nu
}^{a}\right) ^{2}\right)  \nonumber \\
&+& \int d^4x\left(\frac{\lambda^{abcd}}{16}\left( \overline{B}%
_{\mu\nu}^{a}B_{\mu\nu}^{b}-\overline{G}_{\mu\nu}^{a}G_{\mu\nu}^{b}\right)%
\left( \overline{B}_{\rho\sigma}^{c}B_{\rho\sigma}^{d}-\overline{G}%
_{\rho\sigma}^{c}G_{\rho\sigma}^{d}\right)\right)-\int d^4x\left(\frac{9}{4}%
\zeta m^4\right)\;.
\end{eqnarray}
It is easy to check that $\widetilde{\Sigma }_{\mathrm{ph}}$ is not BRST
invariant w.r.t. \eqref{brst}. In fact, it turns out that
\begin{equation}
s\widetilde{\Sigma }_{\mathrm{ph}}=\frac{im}{4}\int d^{4}xG_{\mu \nu
}^{a}F_{\mu \nu }^{a}-\lambda_3\frac{m^2}{16}\int d^4x\left(\overline{B}%
_{\mu\nu}^a-B_{\mu\nu}^a\right)G_{\mu\nu}^a\;.  \label{n2}
\end{equation}
\noindent This equation shows that the breaking of the BRST symmetry %
\eqref{brst} is not linear in the quantum fields, and hence the breaking
terms have to be treated as composite operators \cite{Piguet:1995er}.
Therefore, equation (\ref{n2}) cannot be renormalized as it stands. The two
breaking terms have to be taken into proper account. This is precisely
achieved by introducing the local sources $\left( V_{i\mu \nu },\overline{V}%
_{i\mu \nu },U_{i\mu \nu },\overline{U}_{i\mu \nu }\right) $. In other
words, these sources allow us to take into account e.g. the presence of $%
\int d^{4}xG_{\mu \nu }^{a}F_{\mu \nu }^{a}$ and its renormalization, which
is expressed by the renormalization factor of the source $\overline{U}_{i\mu
\nu }$.

\noindent We would like to understand what happens to the BRST\ symmetry at
the quantum level, when the sources attain their physical values. It is
instructive to study this limit by means of the Slavnov-Taylor identity (\ref
{a5}). Let $\widetilde{\Gamma} _{\mathrm{ph}}$ be the 1PI functional
obtained from $\widetilde{\Gamma} $ when the sources $\left( V_{i\mu \nu },%
\overline{V}_{i\mu \nu },U_{i\mu \nu },\overline{U}_{i\mu \nu }\right) $
attain their physical values
\begin{equation}
\widetilde{\Gamma} _{\mathrm{ph}}=\left. \widetilde{\Gamma} \right| _{%
\mathrm{physical\;value\;of\,\;}\left( V_{i\mu \nu },\overline{V}_{i\mu \nu
},U_{i\mu \nu },\overline{U}_{i\mu \nu }\right) }\;.  \label{a7}
\end{equation}
We can write
\begin{eqnarray}
\left. \int d^{4}x\overline{V}_{i\mu \nu }\frac{\delta \widetilde{\Gamma} }{%
\delta \overline{U}_{i\mu \nu }}\right| _{\mathrm{physical \;value}}&=&-%
\frac{im}{4}\left[\int d^{4}xG_{\mu \nu }^{a}F_{\mu \nu }^{a}\cdot\widetilde{%
\Gamma}\right] _{\mathrm{physical \;value}}  \nonumber \\
&&+\lambda_3\frac{m^2}{16}\left[\int d^4x\left(\overline{B}%
_{\mu\nu}^a-B_{\mu\nu}^a\right)G_{\mu\nu}^a\cdot\widetilde{\Gamma}\right] _{%
\mathrm{physical \;value}} \;,  \label{a8}
\end{eqnarray}
where e.g. $\left[ \left( \int d^{4}xG_{\mu \nu }^{a}F_{\mu \nu }^{a}\right)
\cdot \widetilde{\Gamma} \right] $ stands for the generator of the 1PI Green
functions with the insertion of the composite operator $\left( \int
d^{4}xG_{\mu \nu }^{a}F_{\mu \nu }^{a}\right) $. Of course, it holds that $%
\left[\ldots\cdot\widetilde{\Gamma}\right]_{\mathrm{physical\;value}%
}=\left[\ldots\cdot\widetilde{\Gamma}_\mathrm{ph}\right]$. It follows that
the quantum action $\widetilde{\Gamma} _{\mathrm{ph}}$ obeys the broken
Slavnov-Taylor identity
\begin{equation}
\widetilde{\mathcal{S}}(\widetilde{\Gamma} _{\mathrm{ph}})= \frac{im}{4}%
\left[\int d^{4}xG_{\mu \nu }^{a}F_{\mu \nu }^{a}\cdot\widetilde{\Gamma}%
\right] _{\mathrm{physical \;value}}-\lambda_3\frac{m^2}{16}\left[\int
d^4x\left(\overline{B}_{\mu\nu}^a-B_{\mu\nu}^a\right)G_{\mu\nu}^a\cdot%
\widetilde{\Gamma}\right] _{\mathrm{physical \;value}}\;,  \label{a9}
\end{equation}
where
\begin{eqnarray}
\widetilde{\mathcal{S}}(\widetilde{\Gamma} _{\mathrm{ph}}) &=&\int
d^{4}x\left[ \frac{\delta \widetilde{\Gamma} _{\mathrm{ph}}}{\delta \Omega
_{\mu }^{a}}\frac{\delta \widetilde{\Gamma} _{\mathrm{ph}}}{\delta A_{\mu
}^{a}}+\frac{\delta \widetilde{\Gamma} _{\mathrm{ph}}}{\delta L^{a}}\frac{%
\delta \widetilde{\Gamma} _{\mathrm{ph}}}{\delta c^{a}}+b^{a}\frac{\delta
\widetilde{\Gamma} _{\mathrm{ph}}}{\delta \overline{c}^{a}}+\left( \frac{%
\delta \widetilde{\Gamma} _{\mathrm{ph}}}{\delta \overline{Y}_{i}^{a}}%
+G_{i}^{a}\right) \frac{\delta \widetilde{\Gamma} _{\mathrm{ph}}}{\delta
B_{i}^{a}}+\frac{\delta \widetilde{\Gamma} _{\mathrm{ph}}}{\delta Y_{i}^{a}}%
\frac{\delta \widetilde{\Gamma} _{\mathrm{ph}}}{\delta \overline{B}_{i}^{a}}%
\right.  \nonumber \\
&&\left. +\frac{\delta \widetilde{\Gamma} _{\mathrm{ph}} }{\delta \overline{X%
}_{i}^{a}}\frac{\delta \widetilde{\Gamma} _{\mathrm{ph}}}{\delta {G}_{i}^{a}}%
+\left( \frac{\delta \widetilde{\Gamma} _{\mathrm{ph}}}{\delta X_{i}^{a}}+%
\overline{B}_{i}^{a}\right) \frac{\delta \widetilde{\Gamma} _{\mathrm{ph}}}{%
\delta \overline{G}_{i}^{a}}-\overline{Y}_{i}^{a}\frac{\delta \widetilde{%
\Gamma} _{\mathrm{ph}}}{\delta \overline{X}_{i}^{a}}+X_{i}^{a}\frac{\delta
\widetilde{\Gamma} _{\mathrm{ph}}}{\delta Y_{i}^{a}}\right] \;.  \label{a10}
\end{eqnarray}
\noindent It is worth underlining here that the equation (\ref{a9}) is in
fact nothing more than a direct consequence of the Slavnov-Taylor identity (%
\ref{a5}), when the local sources $\left( V_{i\mu \nu },\overline{V}_{i\mu
\nu },U_{i\mu \nu },\overline{U}_{i\mu \nu }\right) $ attain their physical
values \eqref{phys2}.

\noindent We conclude that $\Gamma _{\mathrm{ph}}$ does not obey an exact
Slavnov-Taylor identity. Of course, (\ref{a9}) translates at the quantum
level the fact that the classical action $\widetilde{\Sigma }_{\mathrm{ph}},$
obtained from $\widetilde{\Sigma }$ by bringing the sources to their
physical values, is not BRST invariant, according to (\ref{n2}). However,
even if $\Gamma _{\mathrm{ph}}$ does not obey an exact Slavnov-Taylor
identity, (\ref{a9}) has far reaching consequences on the behavior of the
1PI Green functions obtained from $\Gamma _{\mathrm{ph}}$, i.e. when the
sources are set to their physical values.

\noindent Let us consider the breaking term $\left[ \left( \int
d^{4}xG_{\mu \nu }^{a}F_{\mu \nu }^{a}\right) \cdot \Gamma \right]
$. Typically, Slavnov-Taylor identities at the level of Green
functions are obtained by acting with a test operator like
$\frac{\delta ^{n}}{\delta \varphi (x_{1})\ldots \delta \varphi
(x_{n})}$, with $\varphi $ any generic field, on expression
\eqref{a9}, and setting all sources and fields equal to zero at the
end. The condition to be fulfilled so that the breaking would be
harmless is quite easily found, since the breaking term will vanish
whenever
\begin{eqnarray}\label{a13}
\frac{\delta ^{n}\left[ \left( \int d^{4}xG_{\mu \nu }^{a}F_{\mu \nu
}^{a}\right) \cdot \Gamma \right] }{\delta \phi (x_{1})\ldots \delta
\phi (x_{n})}=\left\langle \left( \int d^{4}xG_{\mu \nu
}^{a}(x)F_{\mu \nu
}^{a}(x)\right) \varphi (x_{1})\ldots \varphi (x_{n})\right\rangle _{\mathrm{%
1PI}}=0\;,
\end{eqnarray}
meaning that the 1PI Green function with the insertion of the operator $%
\left( \int d^{4}xG_{\mu \nu }^{a}F_{\mu \nu }^{a}\right) $ and with
$n$ amputated external $\phi $-legs should vanish. Thus, if the
condition (\ref{a13}) holds and an analogous one for the other
breaking term, the right hand side of \eqref{a9} is harmless, so
that everything goes as if the theory would fulfill an unbroken
Slavnov-Taylor identity, namely
\begin{eqnarray} \label{a14}\widetilde{\mathcal{S}}(\widetilde{\Gamma
}_{\mathrm{ph}})=0\;.
\end{eqnarray}
\noindent The set of identities for which this happens is quite large.
Certainly, it contains all Slavnov-Taylor identities which are obtained from %
\eqref{a9} by acting only on the original Yang-Mills fields or even the $%
B_{\mu \nu }^{a}$ and $\overline{B}_{\mu \nu }^{a}$ fields. In this case,
there is no way to obtain a nonvanishing contribution to the breaking term
because of the presence of the $G_{\mu \nu }^{a}$-ghost field in the right
hand side of \eqref{a13}. For example, the Slavnov-Taylor identity for the
1PI gluon propagator can be obtained from \eqref{a9} by acting on it with
the test operator $\frac{\delta ^{2}}{\delta c(x)\delta A_{\nu }(y)}$ and
setting all fields and other external sources equal to zero. The breaking
terms will be irrelevant as the Green function $\left\langle \left( \int
d^{4}zG_{\mu \nu }^{a}(z)F_{\mu \nu }^{a}(z)\right) c(x)A_{\nu
}(y)\right\rangle $ as well as the other one are trivially zero.

\noindent We conclude that most Green functions will behave as if
the theory obeys the unbroken Slavnov-Taylor identity \eqref{a14}.
The same considerations outlined for the Slavnov-Taylor identity can
be repeated for the other Ward identities. The corresponding
breaking terms
will always contain the integrated ghost fields $G_{\mu \nu }^{a}$ and/or $%
\overline{G}_{\mu \nu }^{a}$ which, in most cases, will lead to
vanishing contributions when inserted into a Green function, thereby
making the breaking harmless. The beauty in all this is exactly the
fact that the breaking of the Slavnov-Taylor and other Ward
identities can be brought under control at the quantum level by the
introduction of a suitable set of local sources. All the
renormalization results of the action with arbitrary values of the
sources are then preserved once the sources are put equal to
specific values.

\noindent Since the classical part of the action \eqref{finalzondergribov},
obtained by skipping the gauge fixing term \eqref{sgf}, is gauge invariant
w.r.t. the gauge transformations \eqref{gtm2}, we expect that there should
be a nilpotent BRST generator at the quantum level for the gauge fixed
action \eqref{finalzondergribov}. Nevertheless, we have just seen that the
BRST operator \eqref{brst} no longer generates an exact symmetry of the
action \eqref{finalzondergribov}. As it was already discussed in \cite
{Capri:2006ne}, the following nilpotent transformation
\begin{eqnarray}
s^{\prime}A_{\mu }^{a} &=&-D_{\mu }^{ab}c ^{b}\;,  \nonumber \\
s^{\prime}c^{a} &=&\frac{g}{2}f^{abc}c^ac ^{b}\;,  \nonumber \\
s^{\prime}B_{\mu \nu }^{a} &=&gf^{abc}c ^{b}B_{\mu \nu }^{c}\;,  \nonumber \\
s^{\prime}\overline{B}_{\mu \nu }^{a} &=&gf^{abc}c ^{b}\overline{B}_{\mu \nu
}^{c}\;,  \nonumber \\
s^{\prime}G_{\mu \nu }^{a} &=&gf^{abc}c ^{b}G_{\mu \nu }^{c}\;,  \nonumber \\
s^{\prime}\overline{G}_{\mu \nu }^{a} &=&gf^{abc}c ^{b}\overline{G}_{\mu \nu
}^{c}\;,  \nonumber \\
s^{\prime}\overline{c}^{a} &=&b^a\;,  \nonumber \\
s^{\prime}b^{a} &=&0\,,  \nonumber  \label{brstbis} \\
{s^{\prime}}^{2} &=&0\,.
\end{eqnarray}
generates an invariance of \eqref{finalzondergribov}. One shall easily
recognize that there is an intimate connection between the transformations $%
s $ \eqref{brst}, $s^{\prime}$ \eqref{brstbis} and $\delta_s$ \eqref{ss},
namely we have
\begin{equation}  \label{connectie}
s=s^{\prime}+\delta_s\;.
\end{equation}
Then we can say that the breaking of the BRST symmetry $s$ and its
associated Slavnov-Taylor identity is entirely due to the loss of the
supersymmetry $\delta_s$ when the physical limit \eqref{phys2} is taken \cite
{Capri:2006ne}.

\noindent Evidently, the unbroken BRST symmetry $s^{\prime}$ can be used to
construct unbroken Slavnov-Taylor identities between the Green functions of
the massive gauge model \eqref{finalzondergribov}.

\subsection{The case of the massive gauge model with Gribov restriction}

A very similar analysis can be made when we consider the full action %
\eqref{final}. In this case, there are additional breaking terms coming from
the physical limit \eqref{phys1} of the Zwanziger sources. In particular,
applying the transformations $s$ or $s^{\prime }$ on \eqref{final}, we find
\begin{eqnarray}
sS_{\mathrm{physical}} &=&\frac{im}{4}\int d^{4}xG_{\mu \nu }^{a}F_{\mu \nu
}^{a}-\lambda _{3}\frac{m^{2}}{16}\int d^{4}x\left( \overline{B}_{\mu \nu
}^{a}-B_{\mu \nu }^{a}\right) G_{\mu \nu }^{a}  \nonumber \\
&+&\gamma ^{2}\int d^{4}x\left( -gf^{abc}D_{\mu }^{bd}c^{d}\phi _{\mu
}^{ac}+gf^{abc}A_{\mu }^{b}\omega _{\mu }^{ac}+gf^{abc}D_{\mu }^{bd}c^{d}%
\overline{\phi }_{\mu }^{ac}\right) \;,
\end{eqnarray}
or
\[
s^{\prime }S_{\mathrm{physical}}=\gamma ^{2}\int d^{4}x\left(
-gf^{abc}D_{\mu }^{bd}c^{d}\phi _{\mu }^{ac}+gf^{abc}A_{\mu }^{b}\omega
_{\mu }^{ac}+gf^{abc}D_{\mu }^{bd}c^{d}\overline{\phi }_{\mu }^{ac}\right)
\;.
\]
Irrespective of the choice of BRST transformation $s$ or $s^{\prime
}$, the physical Gribov-Zwanziger action is not BRST invariant
anymore. However, repeating the same argument given in the previous
subsection, it turns out that the breaking terms can be treated
consistently at the quantum level, leading to a renormalized broken
Slavnov-Taylor identity. Furthermore, in most cases, everything will
go as if the theory obeys an unbroken Slavnov-Taylor identity,
because the breaking terms are in fact harmless.

\sect{Conclusions}

In this paper, we have shown that the nonlocal gauge invariant operator $%
\mathrm{Tr}\int d^{4}xF_{\mu \nu }(D^{2})^{-1}F_{\mu \nu }$ can be
coupled to the Gribov-Zwanziger action in a localized form. By
embedding this model into a larger class of models with local
sources, we established a comprehensive set of Ward identities,
which were sufficient to prove the renormalizability to all orders
of perturbation theory. Specializing thus to a particular values of
the local sources, we conclude that we have constructed a
renormalizable action \eqref{final} that allows us to study the
gauge invariant mass $m$ in combination with the restriction to the
first Gribov horizon, obtained when the
effective action is minimized w.r.t. the Gribov mass parameter $\gamma ^{2}$%
. This restriction gives a first source of nonperturbative effects in gauge
theories, as explained in the introduction: the gluon/ghost propagator gets
infrared suppressed/enhanced, while the Gribov mass $\gamma $ is fixed in
terms of the QCD scale $\Lambda _{\mathrm{QCD}}$ by means of the requirement
that the effective action is minimized with respect to it.

\noindent We also payed attention to the breaking of the BRST invariance
when the sources are set equal to their physical values. We have elaborated
on the fact that in most cases, the breaking terms in the Slavnov-Taylor
identity are harmless, since they usually induces zero contributions to the
identities between Green functions.

\noindent In the main body of this paper, we have extensively used
and studied the BRST invariance in relation to the
renormalizability. However, there is another major reason why the
BRST symmetry is so important for perturbatively handled gauge
theories. In a certain sense, the BRST invariance is the quantum
version of the gauge invariance, and as such it should play a major
role in reducing the number of relevant (physical) degrees of
freedom, at least at the perturbative level. It is well known that a
BRST symmetry with corresponding nilpotent charge is a very powerful
tool in establishing the unitarity of gauge theories at the quantum
level once a gauge has been chosen, see e.g. \cite
{Kugo:1979gm,Slavnov:1989jh,Frolov:1989az,Henneaux:1992ig,Dudal:2007ch}.

\noindent When we discard the Gribov restriction, we retrieve the
gauge model studied in
\cite{Capri:2005dy,Capri:2006ne,Dudal:2007ch}, enjoying the BRST
symmetry with nilpotent generator \eqref{brstbis}. Therefore, hope
existed that the model might be unitary, i.e. that one would be able
to define a physical subspace $\mathcal{H}_\mathrm{phys}$ of the
total Hilbert state space $\mathcal{H}$, such that
$\mathcal{H}_\mathrm{phys}$ is endowed with a positive norm. This
optimism turned out to be flawed, as it was shown in
\cite{Dudal:2007ch} that the massive gauge model is not unitary.

\noindent In addition to this, the restriction to the Gribov horizon
only makes things worse. First of all, we have lost the (nilpotent)
BRST symmetry, so any potential discussion of the unitarity cannot
be based on BRST related tools. Further, we already mentioned in the
introduction that the Gribov restriction gives rise to an infrared
suppressed gluon propagator, and this suppression is so that the
gluon propagator shows a violation of spectral positivity. Hence,
the gluon is not expected to represent a physical particle, implying
that we should certainly not expect unitarity from the
Gribov-Zwanziger action when the gluons are treated as physical
particles. We should rather expect the opposite. A hint that the
Gribov restriction destabilizes the gluon is also given when we take
a look at the tree level propagator, which in our conventions is
given by \cite{Dudal:2004rx}
\[
\left\langle A_{\mu }^{a}A_{\nu }^{b}\right\rangle _{p}\equiv \delta ^{ab}%
\frac{\mathcal{D}(p^{2})}{p^{2}}\left( \delta _{\mu \nu }-\frac{p_{\mu }{p}%
_{\nu }}{p^{2}}\right) \;,
\]
with the gluon form factor
\begin{equation}
\mathcal{D}(p^{2})=\frac{p^{4}}{p^{4}+2g^{2}N\gamma ^{4}}\;,
\end{equation}
which can also be written in a ``standard'' propagator form
\begin{equation}
\mathcal{D}(p^{2})=\frac{1}{2}\frac{p^2}{p^{2}+i\sqrt{2g^{2}N}\gamma ^{2}}+%
\frac{1}{2}\frac{p^2}{p^{2}-i\sqrt{2g^{2}N}\gamma ^{2}}\;,
\end{equation}
i.e. as the sum of 2 propagators with \emph{imaginary} masses squared.

\noindent A similar reasoning can be applied when we do not
implement the Gribov restriction. As we already outlined in
\cite{Dudal:2007ch}, we can see the massless version of our model
\eqref{finalzondergribov} as an alternative to ordinary Yang-Mills
theory at high energies, based on their equivalence in the
perturbative region \cite{Capri:2006ne}. The benefit of using our
gauge model is that it is possible to couple mass terms $\propto m$
to it without jeopardizing the renormalizability. Then one can start
looking for a sensible gap equation in order to generate a
nonperturbative mass scale $m$. Said otherwise, we could start
looking for a gauge invariant dynamical mass generation mechanism.
The generation of such a mass parameter would break the unitarity at
the level of the gluons, but we must recall that unitarity is only a
prerequisite for the \emph{physical} degrees of freedom. We can
depart our research from the massless (unitary) theory
\cite{Dudal:2007ch}, but we are no longer interested in describing
the perturbative asymptotic high energy regime of QCD, but instead
we are entering a phenomenologically interesting region where e.g.
the gluons already loose their physical meaning as observables. In
this energy region, the gluons should be rather seen as a kind of
quasi particles with a finite lifetime, thus not entering the
asymptotic physical spectrum, which we do not know how to describe.
We can continue to use the gluon propagator etc., albeit the
versions corrected by nonperturbative effects, like an effective
gluon mass and/or Gribov restriction. At first instance, we can
concentrate on the case with only the mass $m$ to be fixed, but at a
later stage, we can also study the influence of the restriction to
the Gribov region $\Omega $, since we have just proven the
renormalizability of this action to all orders.

\section*{Acknowledgments}

The Conselho Nacional de Desenvolvimento Cient\'{i}fico e
Tecnol\'{o}gico (CNPq-Brazil), the SR2-UERJ and the
Coordena{\c{c}}{\~{a}}o de Aperfei{\c{c}}oamento de Pessoal de
N{\'{i}}vel Superior (CAPES) are gratefully acknowledged for
financial support. D.~Dudal is a postdoctoral fellow of the
\emph{Special Research Fund} of Ghent University.

\end{document}